\documentclass[11pt]{article}

\usepackage[utf8]{inputenc}
\usepackage{graphicx}
\usepackage{amsfonts}
\usepackage{amssymb}
\usepackage{amsmath}
\usepackage[mathscr]{eucal}
\usepackage{setspace}
\usepackage{booktabs}
\usepackage[shellescape,latex]{gmp}

\def\href#1#2{#2}   

\def\sqr#1#2{{\vcenter{\hrule height.#2pt
   \hbox{\vrule width.#2pt height#1pt \kern#1pt
      \vrule width.#2pt}
   \hrule height.#2pt}}}

\def\bsqr#1#2{{\vrule width #1pt height#2pt}}
\def\bsquare{{\mathchoice\bsqr66\bsqr66\bsqr33\bsqr33}}
\def\badbreak{\penalty1000}

\newcommand{\cH}{{\cal H}}                             
\newcommand{\cL}{{\cal L}}                    
\def\fir{{\scriptscriptstyle{\text{\rm IR}}}}                 
\def\fa{{\scriptscriptstyle{\text{\rm A}}}}                   
\def\lm0{{\lambda_0}}                                             
\def\v{b}                                                               

\def\Emin{{\langle E_{\scriptscriptstyle{\text{min}}} \rangle}}  
\def\Emax{{\langle E_{\scriptscriptstyle{\text{max}}} \rangle}}  

\newcommand*{\GtrSim}{\smallrel\gtrsim}

\makeatletter
\newcommand*{\smallrel}[2][.8]{%
\mathrel{\mathpalette{\smallrel@{#1}}{#2}}%
}
\newcommand*{\smallrel@}[3]{%
  \sbox0{$#2\vcenter{}$}%
  \dimen@=\ht0 %
  \raise\dimen@\hbox{%
    \scalebox{#1}{%
      \raise-\dimen@\hbox{$#2#3\m@th$}%
    }%
  }%
}
\makeatother

\usepackage{amsmath} 
\usepackage{bm}      

\def\beq{\begin{equation}}
\def\eeq{\end{equation}}
\def\beqs#1\eeqs{\beq\begin{split} #1 \end{split}\eeq}

\long\def\comment#1{}

\usepackage{microtype}
\usepackage[dvipsnames]{xcolor}
\usepackage[colorlinks=true,backref=false, linktocpage=true,
citecolor=PineGreen,urlcolor=PineGreen,linkcolor=PineGreen,pdfpagemode=UseOutlines]{hyperref}

\hypersetup{%
  bookmarksnumbered=true,
  pdftitle = {},
  pdfsubject = {},
  pdfauthor = {},
  pdfkeywords = {}
}

\def\pIR{p}                                          
\def\Wdis{W}                                       

\providecommand{\feast}{\textit{FEAST}}

\begin{document}

\begin{center}
{\Large{\bf Negative-Power Density of States in Chiral-Orthogonal}} \\
\vspace*{.18in}
{\Large{\bf  Disordered Systems and the IR Phase of QCD}} \\
\vspace*{.24in}
{\large{Ji\v{r}\'{\i} Adam, Jr.$^{a}$, Andrei Alexandru$^{b}$, Navdeep Singh Dhindsa$^{c, \star}$, 
Ivan Horv\'ath$^{a,d,e, \dagger}$, \\
\vspace*{.11in}
Frank X.~Lee$^{b}$ and Nilmani Mathur$^{c}$}} \\
\vspace*{.24in}

{\small$^{a}$Theory Department, Nuclear Physics Institute CAS, 25068 $\check{\text{R}}$e$\check{\text{z}}$, 
Czech Republic} \\
\vspace*{.10in}
{\small$^{b}$Department of Physics, The George Washington University, Washington, DC 20052, USA} \\
\vspace*{.10in}
{\small$^{c}$Department of Theoretical Physics, Tata Institute of Fundamental Research, Mumbai 400005, India} \\
\vspace*{.10in}
{\small$^{d}$Department of Physics and Astronomy, University of Kentucky, Lexington, KY 40503, USA} \\
\vspace*{.10in}
{\small$^{e}$Institute of Physics, Slovak Academy of Sciences, 845 11 Bratislava, Slovakia} \\

\vspace*{.15in}
{Nov 10 2026}

\end{center}

\vspace*{0.08in}

\begin{abstract}

The recently discovered IR phase of QCD is characterized by an infrared (IR) sector
of Dirac modes that separates and dynamically decouples from the bulk of the
spectrum, becoming scale invariant in the process~\cite{Alexandru:2019gdm}. 
Among key features of this IR sector is the negative power-law behavior of Dirac 
spectral density near zero with power close to $-1$. Here we start investigating 
to what extent can this behavior be mimicked by chiral Anderson models. In particular,
we study 3-D chiral-orthogonal systems with logarithmic off-diagonal disorder where
negative-power ``spikes'' have been observed~\cite{Evangelou_2003}. We find that
for disorders of moderate strength $\Wdis$, these models in fact do not 
have a power-singular density of states (DOS) at zero energy in the thermodynamic 
limit. Rather, the corresponding power vanishes but the spectrum features 
a very accurate negative-power behavior on an interval around $E \!=\! 1$, 
spreading increasingly closer to $E\!=\!0$ and $E\!=\!\infty$ with increasing $\Wdis$. 
We connect this to the fact that random hoppings $t$ in these quantum systems are 
drawn from a scale-invariant $1/t$ distribution regularized via 
$t \in [e^{-\Wdis/2}, e^{\Wdis/2}]$, and DOS with negative power can naturally 
arise inside that range. In fact, we find that power $p$ in the vicinity of energy $E$
satisfies  $\lim_{\Wdis \to \infty} p(\Wdis, E) = -1$ for all $E$ we access. 
This indicates that, upon removing the cutoff $\Wdis$, quantum mechanics 
transforms the scale-invariant $1/t$ noise into a scale-invariant $1/E$ distribution 
of eigenstates, which mimics the interpretation of scale-invariant Dirac spectral 
densities in the IR phase as being due to scale-invariant glue 
backgrounds~\cite{Alexandru:2019gdm}. Our results imply that, when modeling 
the IR phase, the IR cutoff in chiral Anderson systems needs to be tied to 
their randomness strength $\Wdis$ rather than to their lattice size~$L$.

\renewcommand{\thefootnote}{}
\footnotetext{\hspace*{-.65cm} 
\indent ${}^\dagger${\tt ihorv2@g.uky.edu},${}^\star${\tt navdeep@theory.tifr.res.in}
}
\end{abstract}

\vspace*{0.45in}

\vfill\eject

\noindent
{\bf 1.~Introduction.}
The existence of a qualitatively new thermal regime of strong interactions, now 
referred to as the {\em IR phase}, was first recognized in 
Ref.~\cite{Alexandru:2015fxa}, and the notion was then conceptualized 
in Ref.~\cite{Alexandru:2019gdm} on the basis of key additional lattice evidence 
presented therein. This development has complicated the prevailing view that 
temperature transitions in QCD at zero baryon density are crossovers only, 
a view ushered in by the finding~\cite{Aoki:2006we, Bazavov:2011nk} that 
the chiral transition is of that nature. It has also introduced a new type 
of thermal state accessible in QCD. Indeed, the IR phase transition is 
characterized by the system becoming multicomponent, as the IR sector 
splits away and decouples from the bulk, with the IR component becoming 
scale invariant (at least asymptotically) in the process~\cite{Alexandru:2019gdm}. 
The existence of IR phase is supported by state-of-the-art numerical 
evidence across a range of settings, including thermal QCD with 
$N_f \!=\! 2\!+\!1$ flavors at physical quark 
masses~\cite{Alexandru:2019gdm, Meng:2023nxf, Alexandru:2024tel}. 

\begin{figure}[b]
\begin{center}
    \vskip -0.06in
    \centerline{
    \hskip -0.02in
    \includegraphics[height=4.0truecm,angle=0]{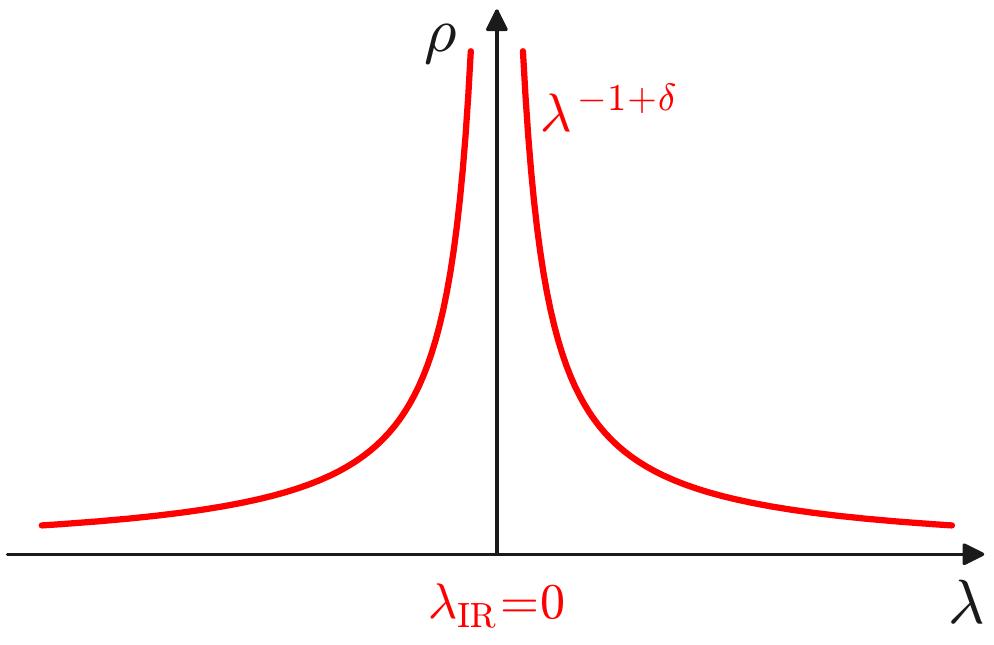}        
    \hskip 0.65in
    \includegraphics[height=4.0truecm,angle=0]{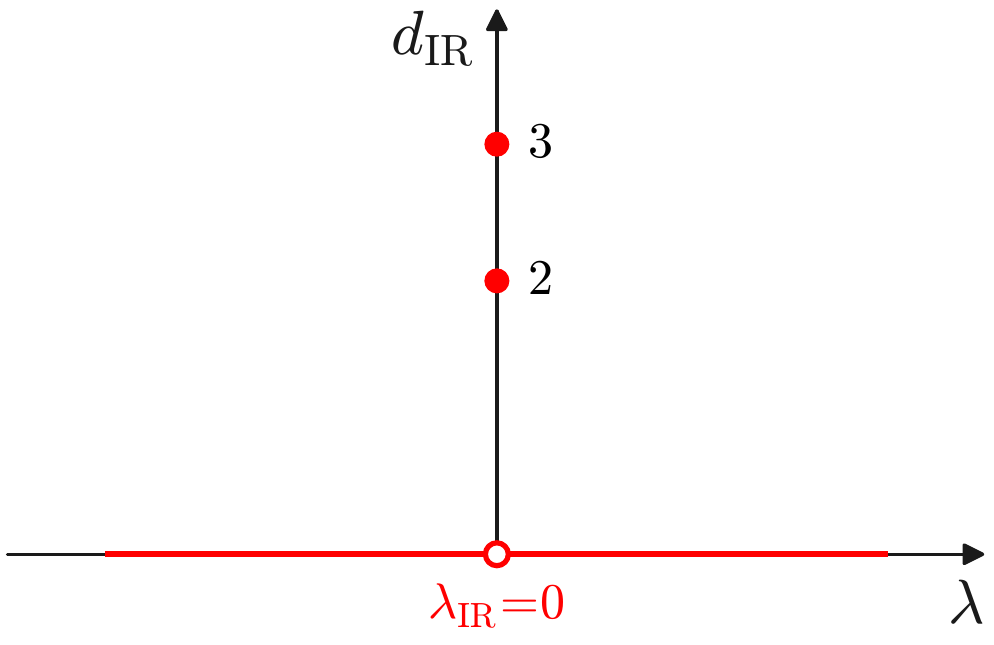}
    }
    \vskip -0.05in
    \caption{Structural fingerprints of the deep-IR (near $\lambda_\fir \!=\!0$) 
    Dirac structure for the system in IR phase.
    Left: negative power law behavior of spectral density.
    Right:  effective spatial dimension of the modes. Note that $d_\fir \!=\! 3$
    for exact zero modes while $d_\fir \!\approx\! 2$ for lowest near-zero modes.
    }
    \label{fig:scheme}
    \vskip -0.40in
\end{center}
\end{figure}

Among the unusual features of the IR transition is that, although it facilitates 
the restoration of a symmetry (scale invariance), this only takes place within 
one part of the resulting multi-component system, the IR component. 
Such a phase transformation has been associated with the emergence of 
a negative power-law behavior of the Dirac spectral density in the IR, 
$\rho(\lambda) \propto \lambda^p$ as $\lambda \to 0$ with 
$p < 0$~\cite{Alexandru:2019gdm}. 
More generally, from the standpoint of IR scale invariance, the phases 
of vectorlike SU(3) gauge theories with fundamental quarks are classified 
via~\cite{Alexandru:2019gdm}
\begin{equation}
     \text{phase}  \;=\;
     \begin{cases} 
        \;\;\text{B}  \quad \text{(IR-broken)}    &     \text{if} \quad  p=0  \\[2pt]
        \;\;\text{IR} \;\; \text{(IR-symmetric)}   &     \text{if} \quad  p<0  \\[2pt]
        \;\;\text{UV} \; \text{(IR-trivial)}            &     \text{if} \quad  p>0        
     \end{cases}
     \quad
     \label{eq:featureA}	     
\end{equation}
where $\pIR = -1 + \delta \, , \; \delta \GtrSim  0$ at least in thermal IR phases 
of pure-glue and real-world QCD. 
Thus, Fig.~\ref{fig:scheme} (left) schematically shows this fingerprint
of the IR phase in the Dirac structure.

An important step toward understanding the IR phase came with the discovery 
of an unusual pattern in the spatial dimensions $d_\fir(\lambda)$ of IR Dirac 
modes~\cite{Alexandru:2021pap, Alexandru:2023xho, Meng:2023nxf}. 
Figure~\ref{fig:scheme} (right) shows the behavior of $d_\fir(\lambda)$ near 
the singular point $\lambda \!=\! \lambda_\fir \!\equiv\! 0$, as inferred from 
the lattice data. The observed collapse of the space occupied by the IR modes 
suggested the involvement of Anderson localization features, and led to 
the metal-to-critical picture of the IR phase 
transition~\cite{Alexandru:2021xoi} (see also the review~\cite{Horvath:2025ypt}).
In the simplest interpretation of this picture, the Anderson-like point 
$\lambda_\fir$, specific to the IR phase, is treated as Anderson mobility edge. 
The scenario also involves another Anderson-like point, $\lambda_\fa \!>\! 0$, 
studied previously in a different 
context~\cite{GarciaGarcia:2006gr, Kovacs:2010wx, Giordano:2013taa, Kehr:2023wrs} 
and likewise identified as a mobility edge. While the scale-invariant deep IR part 
of the Dirac spectrum is governed by $\lambda_\fir$, the buffer regime 
facilitating the separation and decoupling from the bulk is controlled 
by $\lambda_\fa$~\cite{Alexandru:2021pap, Alexandru:2021xoi}.

In this work we start investigating to what extent the two fingerprints of the IR 
phase in Fig.~\ref{fig:scheme} can be reproduced by a suitable Anderson model. 
In other words, we ask how closely, and in what sense, the {\em ``Anderson''} 
and {\em ``Anderson-like''} do really come together for the IR phase. 
The insights gained are expected to provide a more detailed understanding 
of the role randomness plays in generating the thermal states of the IR phase, 
which in turn may help in designing effective models of the phase.
We begin with the negative power law behavior of Dirac spectral density in 
the IR~\cite{Alexandru:2019gdm}, which is most central to the notion of IR 
phase, particularly to its scale invariance. Indeed, the observed near-pure 
power over wide IR range was immediately ascribed to approximate 
statistical self-similarity (IR scale invariance) of the underlying glue 
fields~\cite{Alexandru:2019gdm}.\footnote{The arguments in
Ref.~\cite{Alexandru:2019gdm} were supplemented with more
formal analysis in Ref.~\cite{Horvath:2025jzl}.} 
Modeling this aspect of QCD with a simple Anderson model then casts 
this issue into one of understanding whether and how a self-similar ``field'' 
describing fermion hopping gives rise to a power-law density of 
states in the IR.
%

In Sec.~2 we describe chiral-orthogonal Anderson models with off-diagonal 
disorder, which is the class we study here. Their suitability is motivated 
by arguments of standard nature (chiral symmetry, structure of operators)
and the previous observation of ``spikes'' in DOS~\cite{Evangelou_2003}
for ``logarithmic'' disorder, that could be analogs of deep-IR 
negative-power behavior seen in QCD. However, there is another feature 
of these models we make explicit here (Sec.~3) that is central to our 
purposes stated in the previous paragraph: that ``logarithmic'' 
off-diagonal disorder of strength $W$ entails a random hopping field 
$t_{r,r+i}$ where the individual hoppings are distributed as $1/t$ 
in the interval $[e^{-W/2}, e^{W/2}]$. 
This is indeed essential since the two key features of IR phase 
extracted via lattice QCD, namely the scale invariance in the form 
of self-similarity~\cite{Alexandru:2019gdm} and the importance
of randomness~\cite{Alexandru:2021pap, Alexandru:2021xoi}, are 
combined by the model in a distilled form and taken to the extreme:   
the hopping fields are strictly self-similar 
(in the $W\!\to\!\infty$ limit) and strictly random.\footnote{Strict 
self-similarity stems from the full $1/t$
distribution of noise upon removing the noise cutoff $W$, while the strict
randomness from statistical independence of individual hoppings.}

Given the above, the central objective of the present work is to clarify
how regularized scale-invariant (self-similar) hopping noise translates 
into spectral features of a model Hamiltonian, and what happens 
upon removing the noise regularization. In Secs.~4-6 we take key 
steps toward answering that question in case of $1/t$ noise. In 
particular, we show numerically that the noise-regularization interval
$[e^{-W/2}, e^{W/2}]$ has a clear imprint in the spectrum of the model 
Hamiltonian and that DOS indeed exhibits a power law inside of it, 
with significant violations only toward the edges of the regularization
interval. However, we also show that, at least in some range of finite 
$W$,  the power in the vicinity of $E \!=\! 0$ vanishes in the thermodynamic 
limit, and there is no true negative-power divergence in DOS.
At the same time, the power inside the noise-regularization interval 
approaches $-1$ and spreads over the entire positive real axis upon 
the removal of noise cutoff ($W \!\to\! \infty$). In Sec.~7 we analyze
the consequences of these findings for the possibility of modeling
the scale-invariant IR component of IR phase.



\smallskip
\noindent
{\bf 2.~Chiral-Orthogonal Anderson Models.}
To identify Anderson models mimicking QCD near $\lambda_\fir\!=\!0$, it is 
natural to begin with models having a suitable form of chiral invariance, 
so as to respect the $\lambda \rightarrow -\lambda$ symmetry of the massless 
Dirac operator. Separately, the Dirac operator couples a quark at a given 
space-time point only to its immediate neighborhood (quark hopping), which 
singles out Anderson models with local off-diagonal (hopping) disorder. 
The two aspects come together in a simple manner when chiral 
symmetry is realized via a bipartite nearest-neighbor Hamiltonian. 
On the hypercubic lattice the Hamiltonian reads 
(see, e.g.,~\cite{Evangelou_2003,garcia2006anderson})
\begin{equation}
     H \,= \sum_{r\in \cL, i} t_{r,r+i} \, 
     \bigl( c_r^\dagger c_{r+i} + c_{r+i}^\dagger c_{r} \bigr)  
     \label{eq:008} 
\end{equation}
where $r$ is the position vector on the lattice $\cL$ of $L^D$ sites and 
$i=1,2,\ldots,D$ labels the lattice directions on a flat torus. Note that there is 
a single fermion operator per site. Randomness enters via real hopping 
strengths $t_{r,r+i}$ drawn independently from a common distribution. 
As written, the system is bipartite (chirally symmetric) only for even $L$, 
the two parts being the even and odd sublattice. For odd $L$ it becomes
bipartite by setting all even-even and odd-odd hopping strengths to zero. 
However, here we will assume even $L$ from now on.

Given the Eq.~\eqref{eq:008}, the wave functions $\chi = \chi(r)$ of 1-particle states 
$\mid \chi\, \rangle \!=\!  \sum_r \chi(r) \, c_r^\dagger \mid 0\,\rangle$ are governed 
by the 1-particle Hamiltonian matrix
\begin{equation}
     \cH_{r,r'} = \sum_i t_{r,r+i} \, \delta_{r+i,r'}  \,+\, 
                                   t_{r-i,r} \, \delta_{r-i,r'}
     \label{eq:018}      
\end{equation}
which is real symmetric in any dimension. In this work we will impose 
periodic boundary conditions on the eigenmodes. Notice that the 
$E \rightarrow -E$ symmetry of its spectrum is easily seen. 
Indeed, since for the eigenmode $\chi$ of energy $E$ we have  
\begin{equation}
     ( {\cal H} \chi ) (r) \,=\, 
     \sum_{i} \, t_{r,r+i} \, \chi(r+i) + t_{r-i,r} \, \chi(r-i)  \,=\, E \, \chi(r)
     \label{eq:028}      
\end{equation}
flipping the sign of $\chi$ on even (odd) sublattice but keeping it unchanged 
on odd (even) sublattice produces the eigenmode with energy -E. 

Before introducing the precise nature of noise governing hoppings $t$, it is 
instructive to know the spectrum for $t_{r,r+i} \equiv 1$ which
corresponds to translation invariant non-random system. 
The eigenmodes are the plane waves $\chi(r) \propto \exp(i \, q \cdot r)$,
with periodic boundary conditions selecting out the momenta  
$q_i = (2\pi/L) k_i$ with $k_i = 0,1,\ldots ,L \!-\! 1$. The spectrum reads
\begin{equation}
       E(k_1, \ldots, k_D) \,=\, 2 \sum_{i=1}^D 
       \cos \left( \frac{2\pi}{L} k_i \right)
       \quad\; , \quad\;
       k_i \in \{0,1,\ldots, L-1 \} 
       \label{eq:038}             
\end{equation}
and is contained in the interval $[-2D, 2D]$. 

\smallskip
\noindent
{\bf 3.~Logarithmic Disorder.}
For our purposes, the key feature of $D\!=\!3$ chiral orthogonal Anderson
models on bipartite lattices is the previously reported appearance of strong
spikes in the density of states (DOS) for off-diagonal ``logarithmic disorder''
(see~\cite{Evangelou_2003} and references therein). This technical term
refers to hoppings $t$ whose logarithm $\ln t$ is distributed uniformly on
the interval $[-W/2, W/2]$. A sample is normally obtained by drawing $y$ from
this distribution and setting $t = e^{y}$. But this procedural description may 
obscure the fact that it corresponds to hoppings $t$ distributed as
\begin{equation}
       P(t) = \frac{1}{W}\,\frac{1}{t} \quad, \quad
       t \in \bigl[\, e^{-W/2},\, e^{W/2} \,\bigr]
       \label{eq:048}
\end{equation}
In other words, the Anderson models~\eqref{eq:008} with logarithmic 
disorder describe quantum motion in a random medium generating 
hoppings that follow a scale-invariant $1/t$ distribution,  
{\em regularized} via an explicit cutoff $W$ in a way specified by \eqref{eq:048}.
This is well suited to our goals, since the Dirac operator--Anderson 
Hamiltonian association we follow puts the gauge field and 
the ``hopping field'' into mutual correspondence. Indeed, it is hard 
to think of a simpler natural analog of random self-similar glue 
field regularized on the lattice than the hopping field above.

A few relevant remarks regarding the noise regulator $W$. 
(i) Taking $W \!\to 0$ turns the noise off and corresponds 
to translation-invariant non-random system ($t_{r,r+i} \! \equiv \! 1$) 
with spectrum \eqref{eq:038}, while going the opposite way makes 
the system noisier. The limit $W \!\to\! \infty$ removes the regulator and 
defines the scale-invariant random Anderson system~\eqref{eq:008} via
the associated limits of observables. 
(ii) Since $1/t$ is non-normalizable on the non-negative real axis, its
regularization is necessary. This situation parallels that in lattice QCD, 
which is itself a regularization of QCD. However, setting up 
a correspondence between cutoffs on the Anderson and QCD sides is 
not obvious because, among other things, the explicit cutoff for the noise 
and the lattice size are independent parameters in the Anderson case. 
Resolving related issues is an important aspect of the present investigation.
(iii) While it naively appears that one could consider the noise with 
independent low/high cutoffs $t_1$ and $t_2$, namely the distribution
$P(t) = (\ln t_2/t_1)^{-1}(1/t)$ with $t \in [t_1,t_2]$, this is redundant. 
Indeed, due to the form-invariance of $1/t$ distribution upon the rescaling 
$t \to st$, only distinct ratios $t_2/t_1$ lead to distinct physical 
results.\footnote{Spectra of the Hamiltonian \eqref{eq:018} in backgrounds
with the same ratio $t_2/t_1$ are related by simple rescaling.} In 
the condensed matter literature this ratio was chosen to be parametrized 
by $\exp(W)$ and we follow this convention here.

\begin{figure}[t]
\begin{center}
    \vskip -0.06in
    \centerline{
    \hskip -0.02in
    \includegraphics[height=4.4truecm,angle=0]{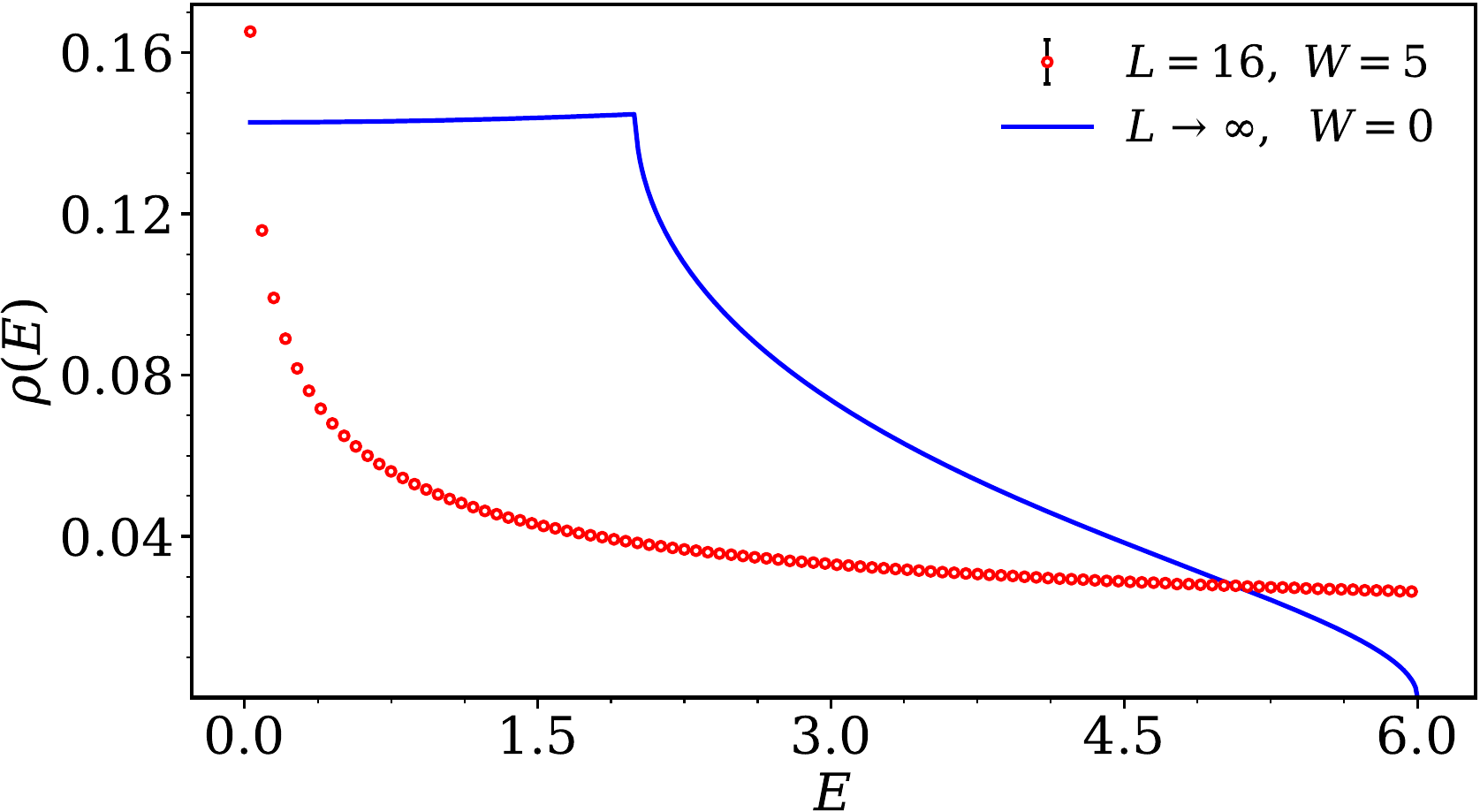}        
    \hskip -0.00in
    \includegraphics[height=4.4truecm,angle=0]{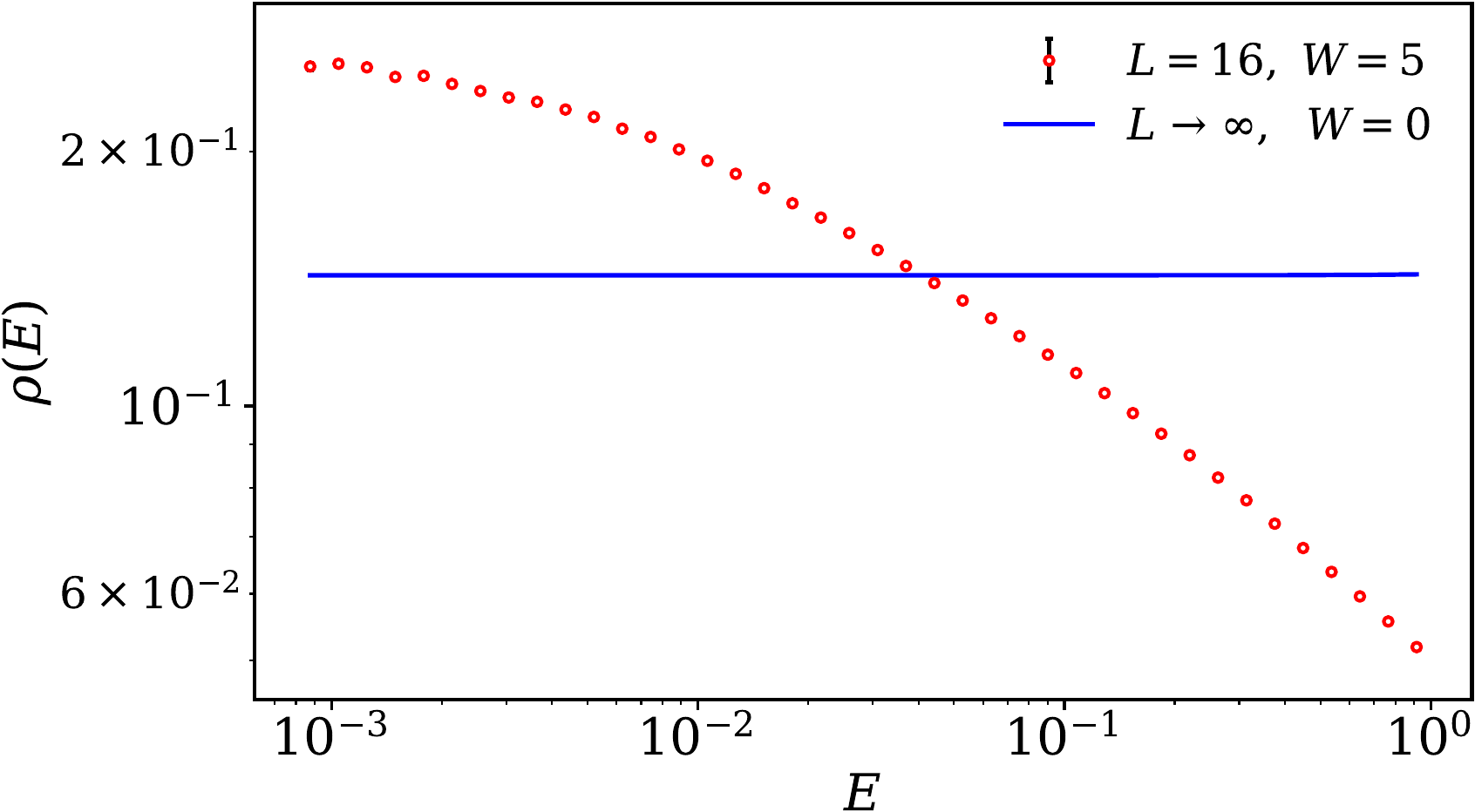}
    }
    \vskip -0.05in
    \caption{Left: density of states for non-random spectrum \eqref{eq:038} in 
    $D \!=\! 3$ over its spectral range (blue line) compared to the case of 
    $W \!=\! 5$ logarithmic noise. 
    Right: zoom (log-log form) to the lowest part of the spectrum where DOS 
    of $L \!=\! 16$ random system doesn't change upon increasing $L$.
    }
    \label{fig:deform}
    \vskip -0.40in
\end{center}
\end{figure}

We close this section by illustrating how turning on randomness  
($W \!=\!0 \,\rightarrow\, W \!>\! 0 $) affects non-random free spectrum 
in terms of spectral density. Considering the relevant 
$D \!=\! 3$ case studied here, Fig.~\ref{fig:deform} (left) shows DOS 
for $W \!=\! 0$ over the extent of its spectrum Eq.~\eqref{eq:038}, and 
compares it to the $W \!=\! 5$ numerical result. Note that, although
the range of noise is severely restricted by the cutoff 
($[e^{-W/2}, e^{W/2}]$ only spans about 2 orders 
of magnitude) the distribution of states over shown energies
drastically changes  in favor of spiky low end. At the same time, 
the log-log detail in deep IR (below $e^{-W/2}$) and the leveling off
the DOS raises interesting relevant questions (see Sec.~5).


\begin{figure}[b]
\begin{center}
    \vskip -0.06in
    \centerline{
    \hskip -0.02in
    \includegraphics[height=4.4truecm,angle=0]{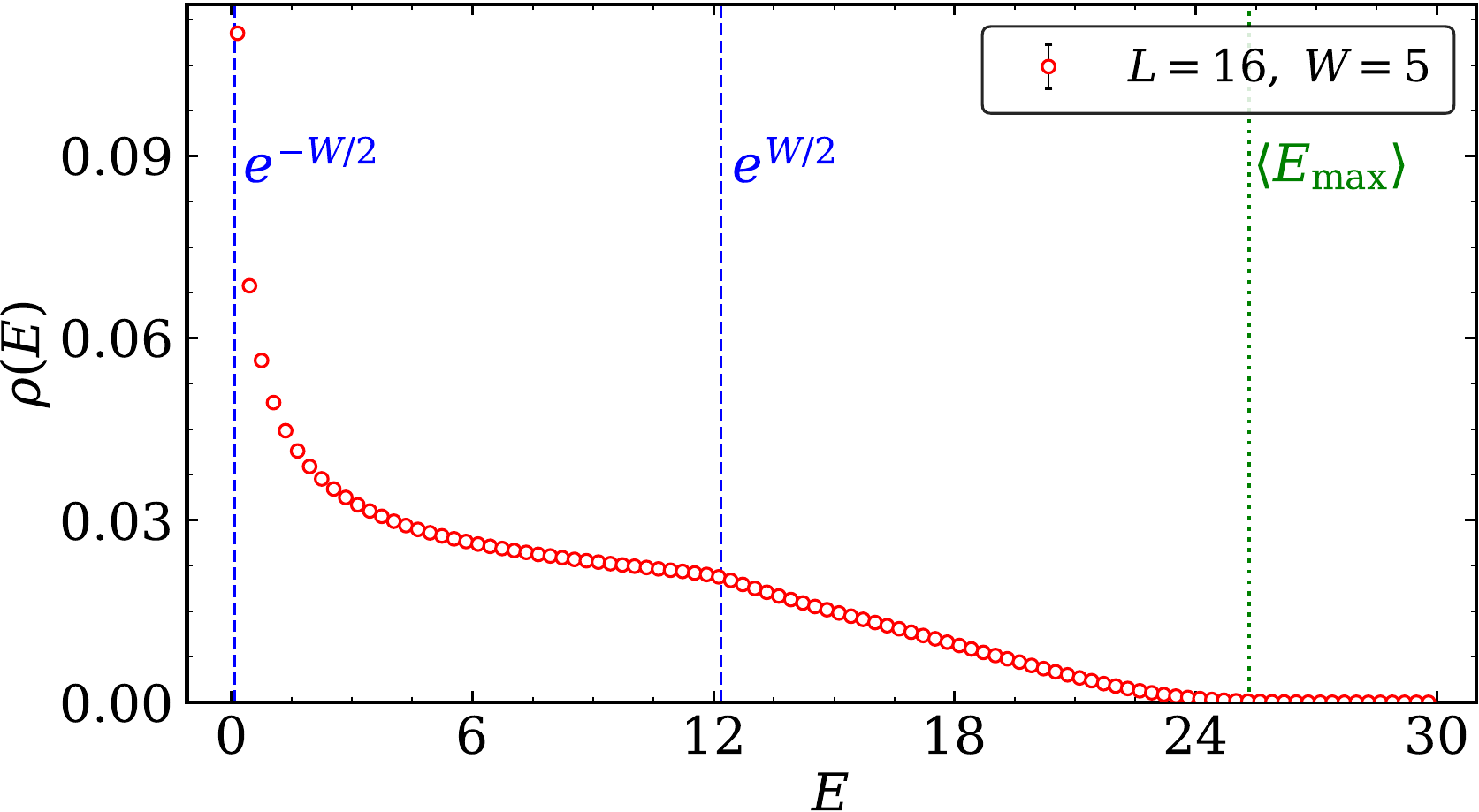}        
    \hskip 0.05in
    \includegraphics[height=4.4truecm,angle=0]{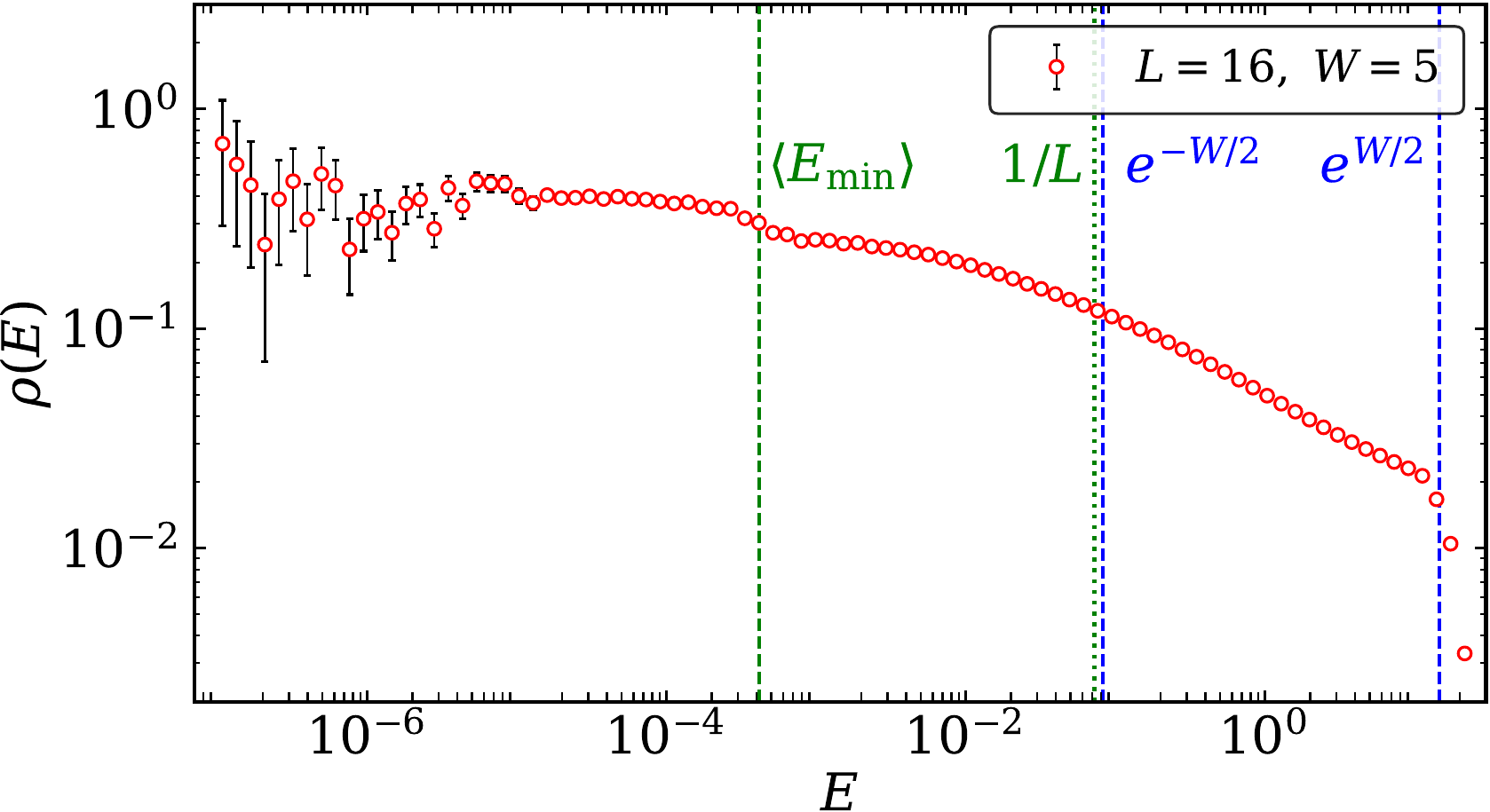}
    }
    \vskip 0.00in
    \caption{Density of states for $W \!=\! 5$, $L \!=\! 16$ system in the lin-lin (left)
    and log-log (right) form over the entire spectral range. Effective spectral bounds 
    $\Emin$ and $\Emax$, as well as all explicit cutoffs are shown. 
    See the discussion in the text.
    }
    \label{fig:full_spectra}
    \vskip -0.40in
\end{center}
\end{figure}

\smallskip
\noindent
{\bf 4.~Basic Spectral Features and the Cutoffs.}
As explained in previous sections, the models we study here are labeled by 
the lattice size $L$ (conventional IR cutoff) and the noise-cutoff parameter $W$. 
To study them numerically, we compute the spectra of the Hamiltonian 
\eqref{eq:018} using the FEAST package 
\cite{polizzi2009density, 7877094, di2016efficient, gavin2017improved, polizzi2020feast}. 
The elementary description of the underlying algorithm, and the particular 
implementation of it we utilize, is described in the Appendix A.

In order to expose its global structure,  we show in Fig.~\ref{fig:full_spectra} 
the DOS, defined in the usual way,  for $W \!=\!5 \,,\, L\!=\!16$ system over its 
full spectrum. The displayed $\rho(E)$ is from $4\times10^4$ ``configurations'' 
of logarithmic noise. 
In the left plot, the low-high noise cutoffs and the average maximal 
eigenvalue $\Emax$ are marked by the blue and green vertical lines respectively. 
Note that the IR peak structure visibly changes its character at the high noise 
cutoff $e^{W/2}$. The log-log plot reveals richer IR structure, and 
fully exposes the special role of the cutoff range $[e^{-W/2}, e^{W/2}]$ 
in the spectrum, with approximately linear behavior (power law in 
lin-lin). The slope gradually changes toward smaller values as 
even deeper IR is approached. The average minimal non-negative 
energy $\Emin$ and its maximal counterpart $\Emax$ delimit 
the effective extent of the spectrum. 

For our analysis of DOS in the following sections it is important that 
the spectral bounds satisfy 
\begin{equation}
      \Emin  \ll\, e^{-W/2}   \qquad \qquad \quad
      e^{W/2} <\, \Emax
      \label{eq:058}        
\end{equation}
for sufficiently large $L$. 
Such $L$-regime not only renders finite-L errors in $\rho(E)$ negligible 
for $E \!\in\! [e^{-W/2}, e^{W/2}]$, but also allows for a controlled 
extraction of its power-law behavior in the vicinity of $E \!=\! 0$
in $L \!\to\! \infty$ limit (Secs.~5,6). 
%
%
We verified that Eq.~\eqref{eq:058} holds in all numerical experiments we 
performed. Moreover, these relations are satisfied even for small $L$. 
For example, they hold for all $5 \le W \!\le 60$ at fixed $L\!=\!16$. 

\begin{figure}[t]
\begin{center}
    \vskip -0.06in
    \centerline{
    \hskip -0.02in
    \includegraphics[height=4.7truecm,angle=0]{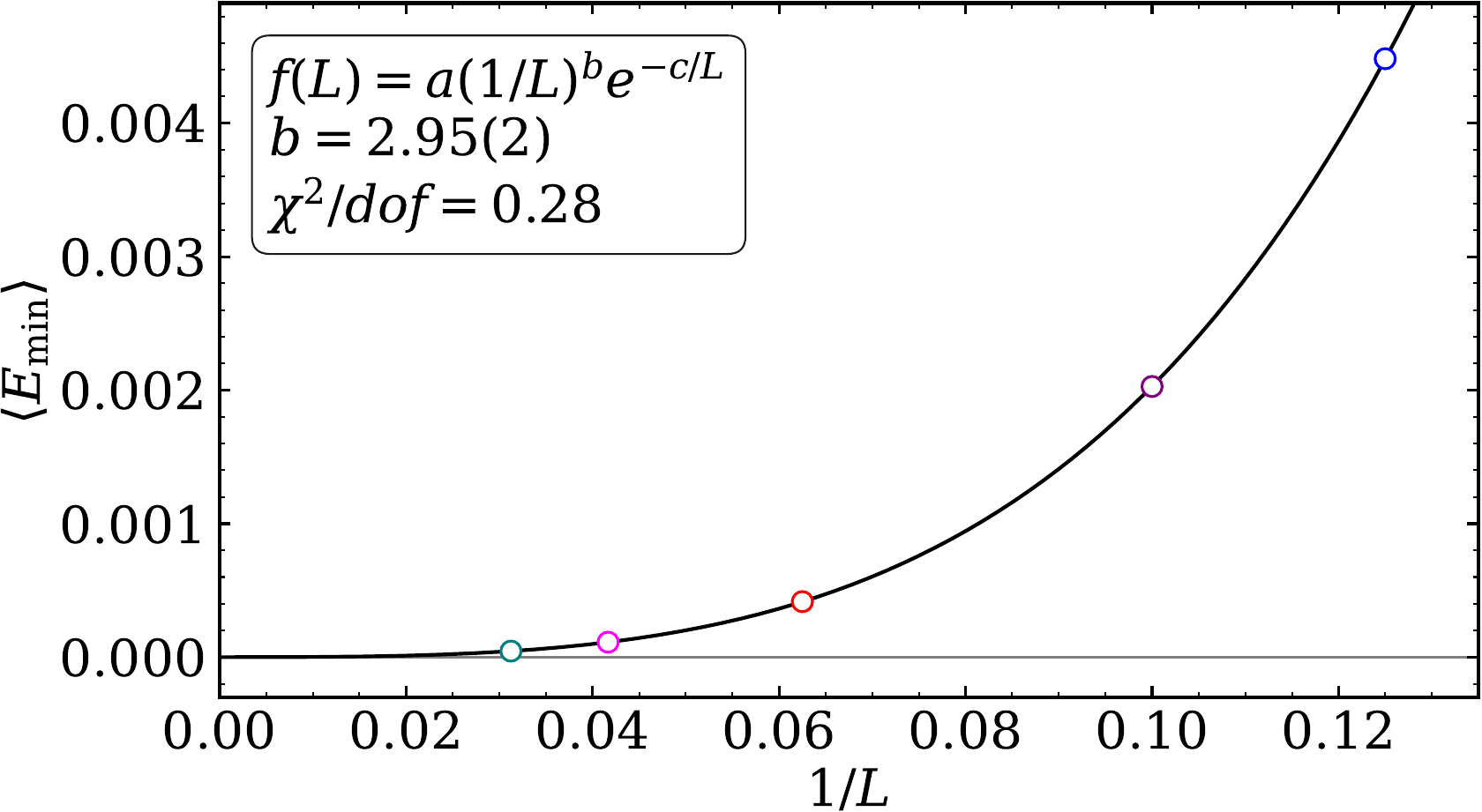}        
    \hskip 0.15in
    \includegraphics[height=4.7truecm,angle=0]{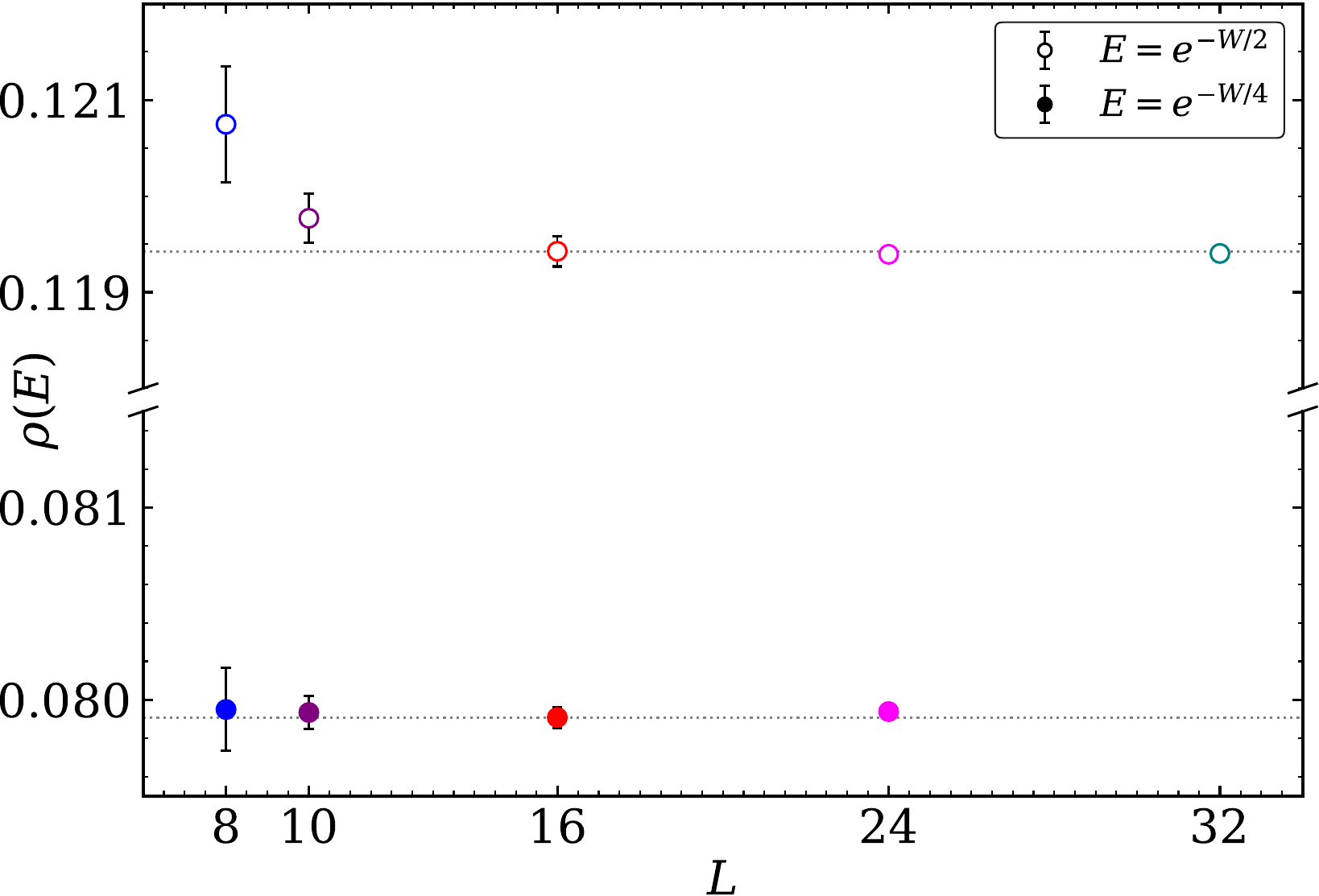}
    }
    \vskip 0.00in
    \caption{Left: the average minimal positive eigenvalue as a function of
    $1/L$ for the system at $W \!=\! 5$. The form of the fit was originally proposed in 
    Ref.~\cite{Meng:2023nxf}. Right: the $L$-dependence of $\rho(e^{-W/2})$ 
    and $\rho(e^{-W/4})$ at $W \!=\! 5$. See the discussion in the text.
    }
    \label{fig:cutoffs_rhoL}
    \vskip -0.40in
\end{center}
\end{figure}

To gain some insight into such behavior, we show in Fig.~\ref{fig:cutoffs_rhoL} 
(left) the $L$-dependence of $\Emin$ at $W \!=\! 5$. 
The decay of $\Emin$ with increasing $L$ is unusually fast, close to $L^{-3}$. 
Furthermore, the right plot in Fig.~\ref{fig:cutoffs_rhoL} shows the associated 
$L$-dependence of 
$\rho(e^{-W/2})$, that is of the density at the lower edge of the relevant 
interval. Although this is where the largest corrections to $L \!\to\! \infty$ limit 
are expected, the value saturates at $L\!=\!16$ already. Note that essentially
no violations are observed for $\rho(e^{-W/4})$. 

\smallskip
\noindent
{\bf 5.~Deep IR at Finite Disorder Strength.}
When speaking of IR asymptotics in field theories, one usually refers to 
the behavior in the vicinity of strict IR limit in energy-like/distance-like 
variable in question.  
For example, by IR power law behavior of Dirac spectral density in QCD
we mean $\lim_{L \to \infty} \rho(\lambda,L) \propto \lambda^p$ for 
$\lambda \rightarrow 0$.  While this is less implicit for models 
of condensed matter physics, it is the case for chiral bipartite models 
due to the $E \!\rightarrow\! -E$ symmetry of the spectrum and 
the ensuing role of $E\!=\!0$. Moreover, if $p \!<\! 0$, then the notion 
of a singular spectrum only makes sense in relation to the vicinity of zero. 
Like in Ref.~\cite{Evangelou_2003}, this is usually not discussed, and a single 
power is implicitly assumed. However, given the special status of the noise 
interval $[e^{-W/2}, e^{W/2}]$ at finite $W$ (Secs.~3,4), we define two, 
possibly distinct powers, namely in the vicinity of $E\!=\!0$ and 
$E \!=\! 1$, i.e.
\begin{equation}
   \lim_{L \to \infty} \rho(E,L,W) \propto E^{\,p_1(W)} \;\, , \;\, E \to 0^+
   \qquad\,  \text{and}  \qquad\,
   \lim_{L \to \infty} \rho(E,L,W) \propto E^{\,p_2(W)} \;\, , \;\, E \to 1 \quad
   \label{eq:068}            
\end{equation}
It is clearly $p_2$ that is directly tied to the regularized scale invariance of the 
underlying randomness. 

\begin{figure}[t]
\begin{center}
    \vskip -0.06in
    \centerline{
    \hskip -0.02in
    \includegraphics[height=4.4truecm,angle=0]{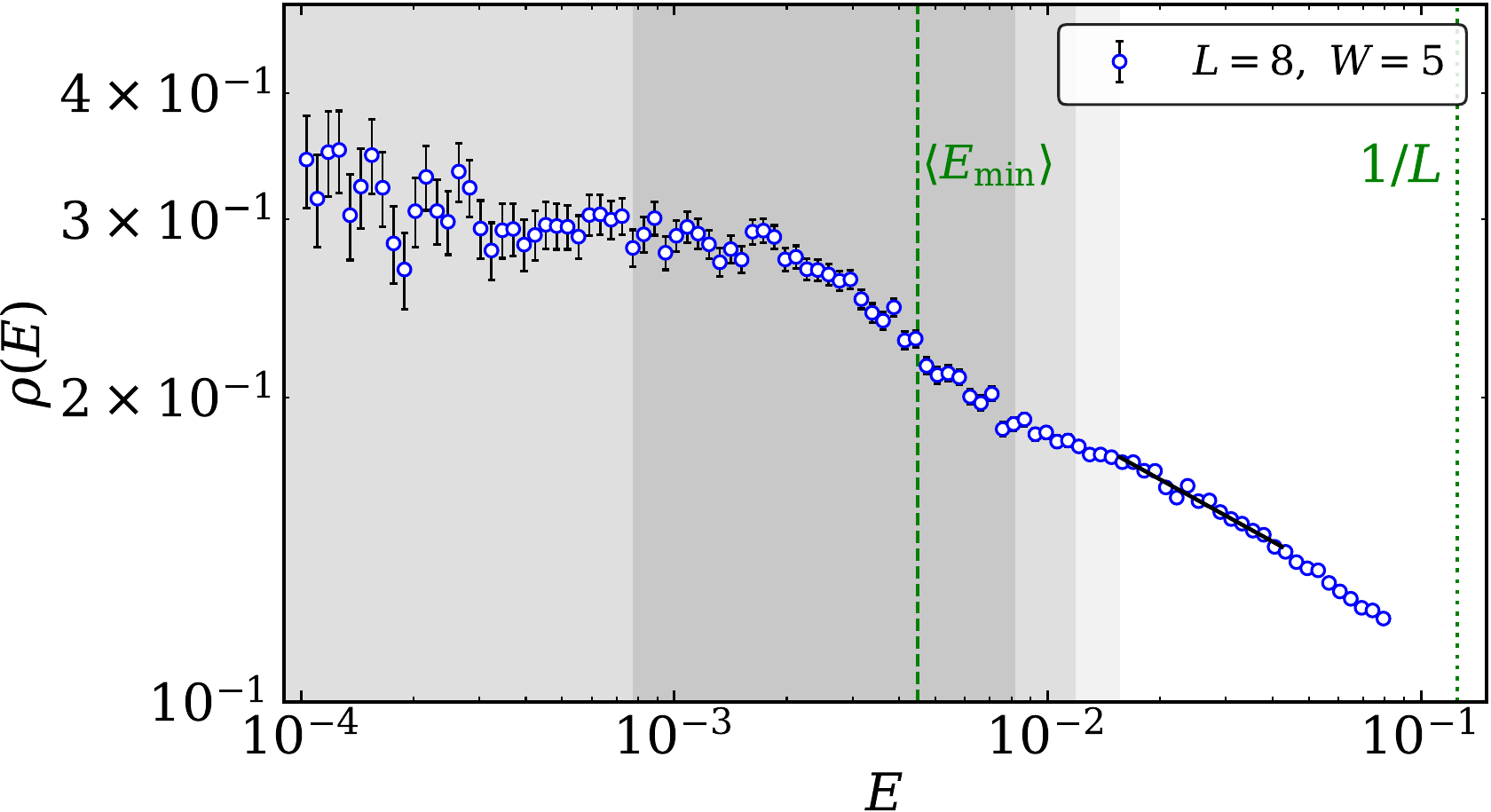}        
    \hskip 0.13in
    \includegraphics[height=4.4truecm,angle=0]{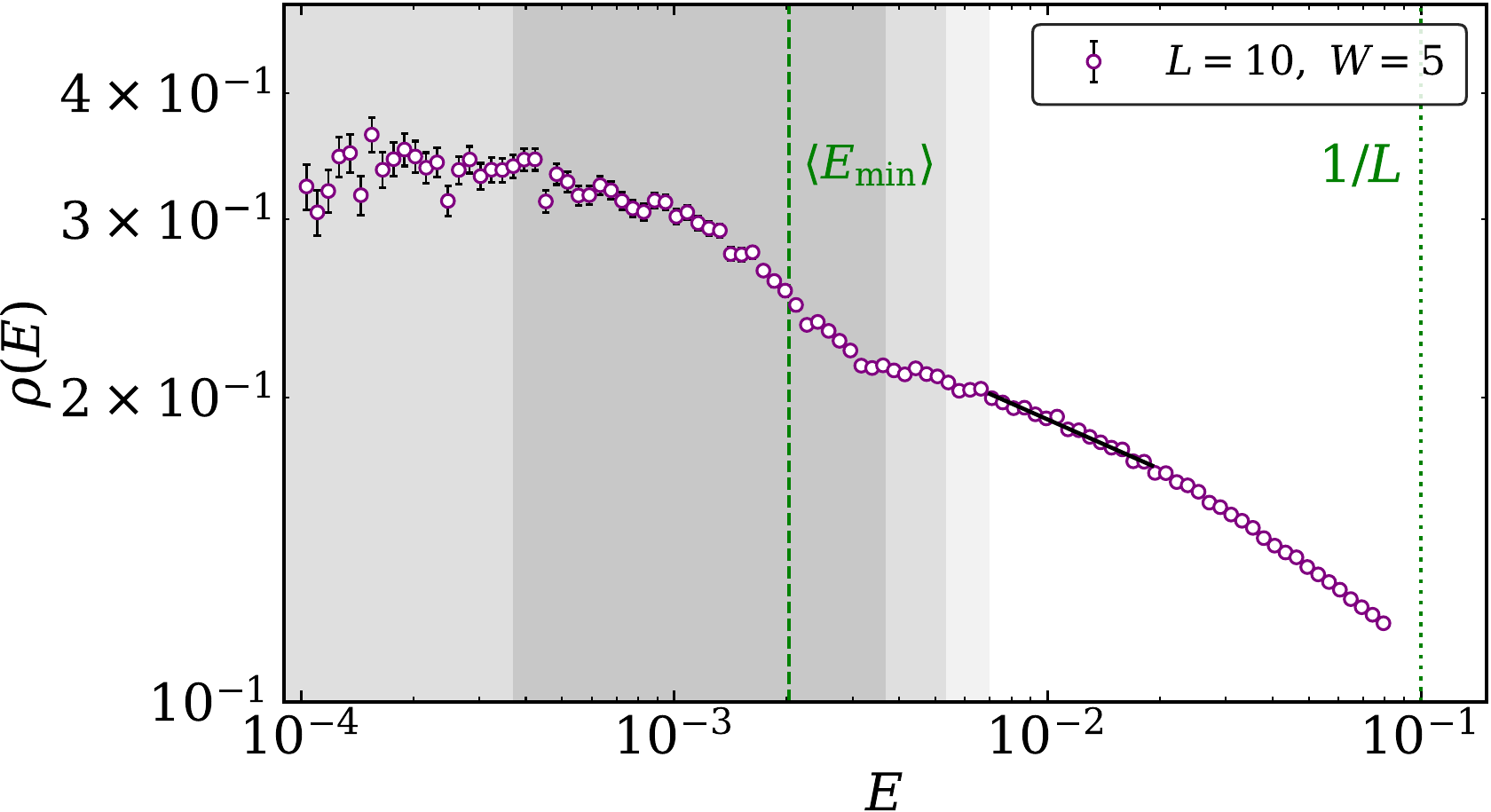}
    }
    \vskip 0.13in
    \centerline{
    \hskip -0.02in
    \includegraphics[height=4.4truecm,angle=0]{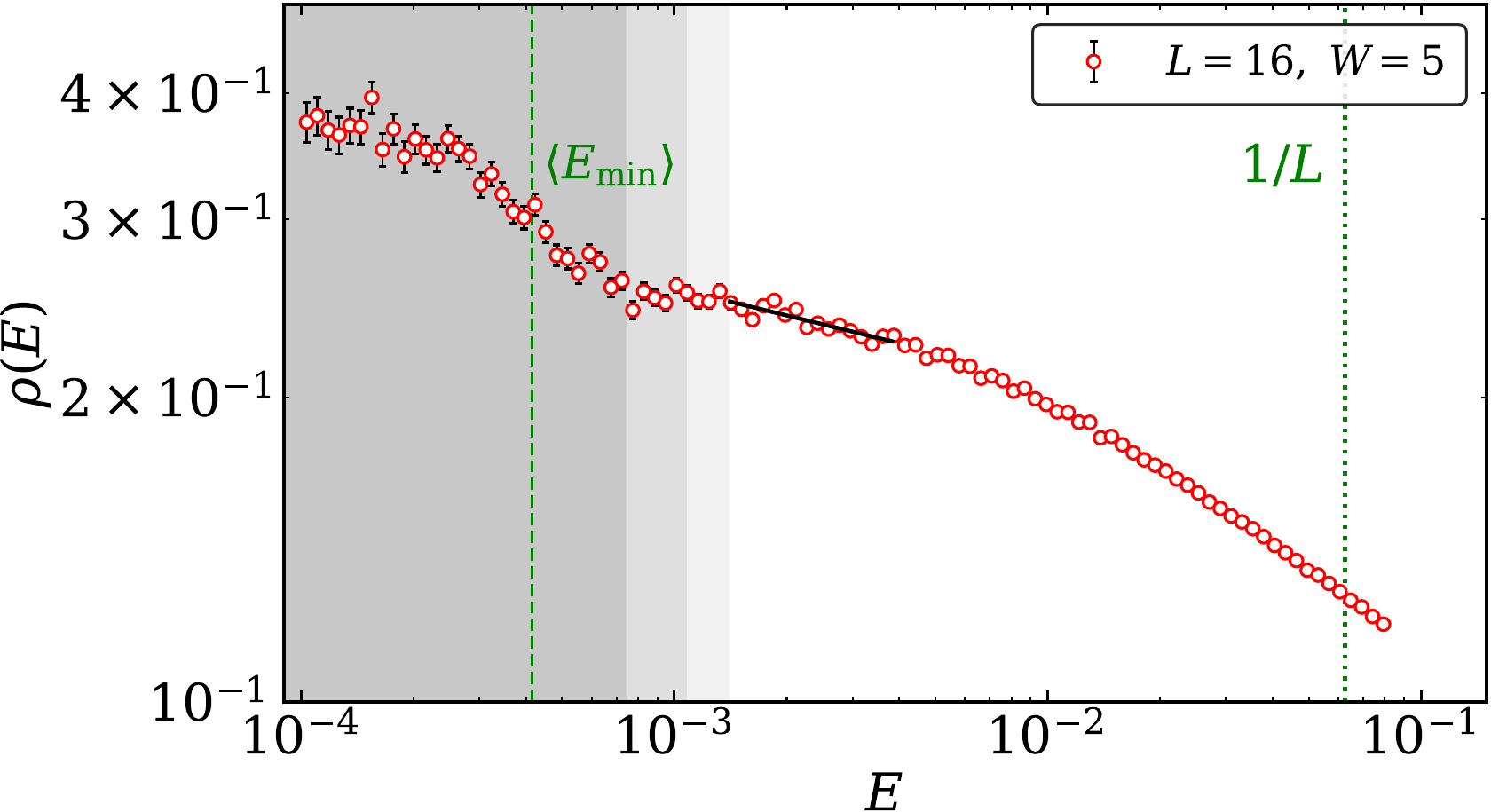}        
    \hskip 0.13in
    \includegraphics[height=4.4truecm,angle=0]{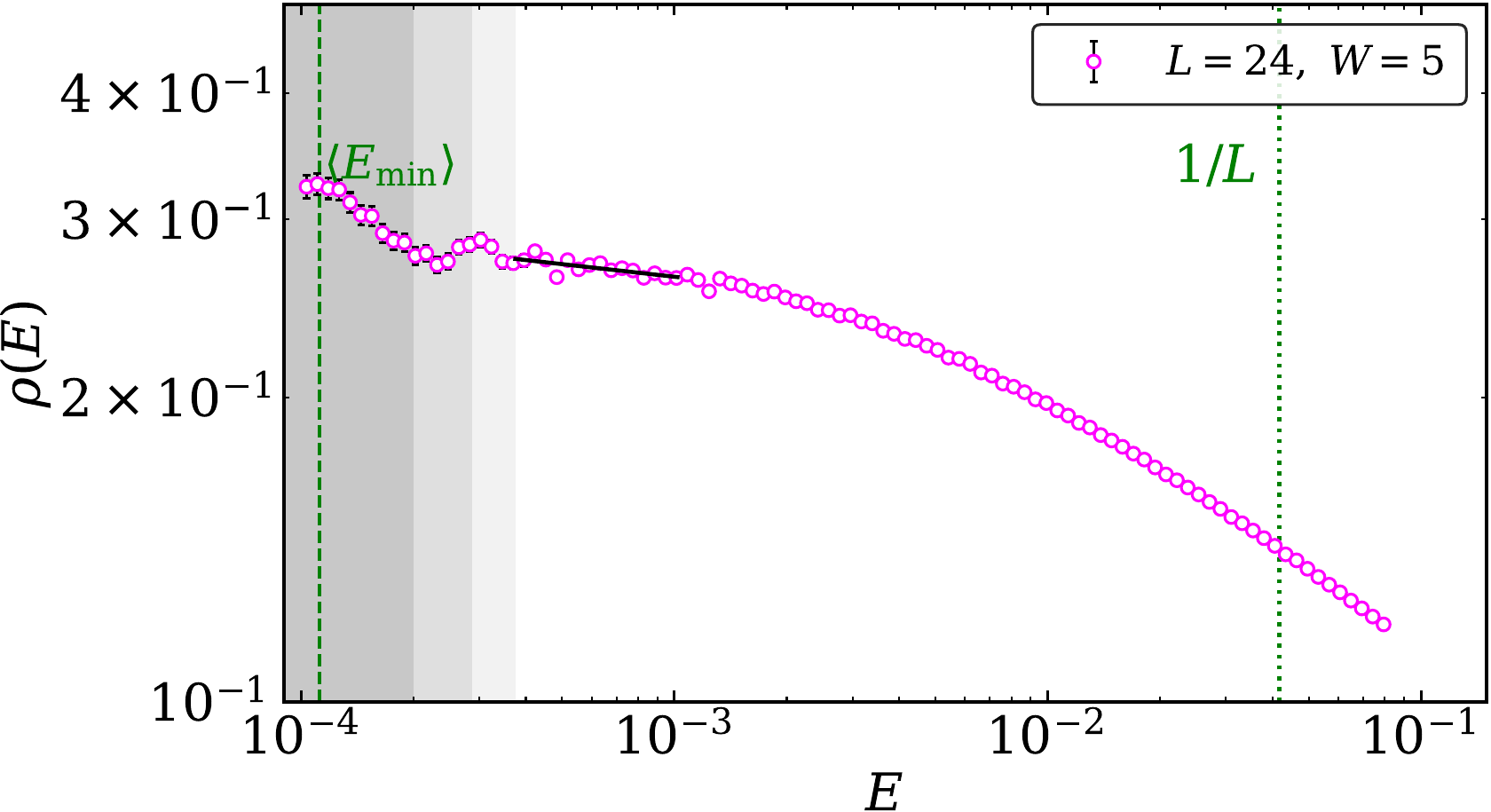}
    }
    \vskip 0.13in
    \centerline{
    \hskip -0.02in
    \includegraphics[height=4.4truecm,angle=0]{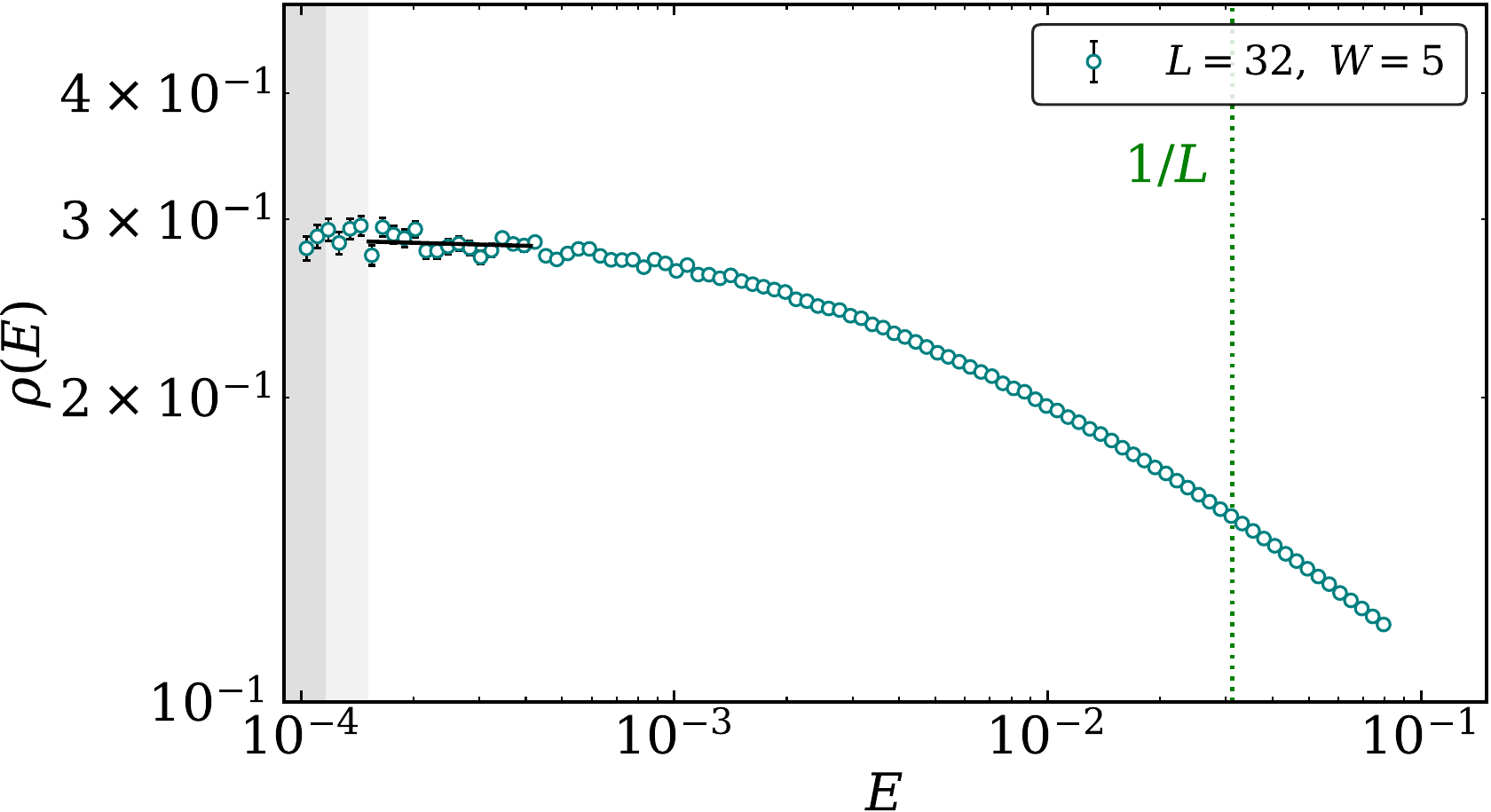}        
    \hskip 0.17in
    \includegraphics[height=4.4truecm,angle=0]{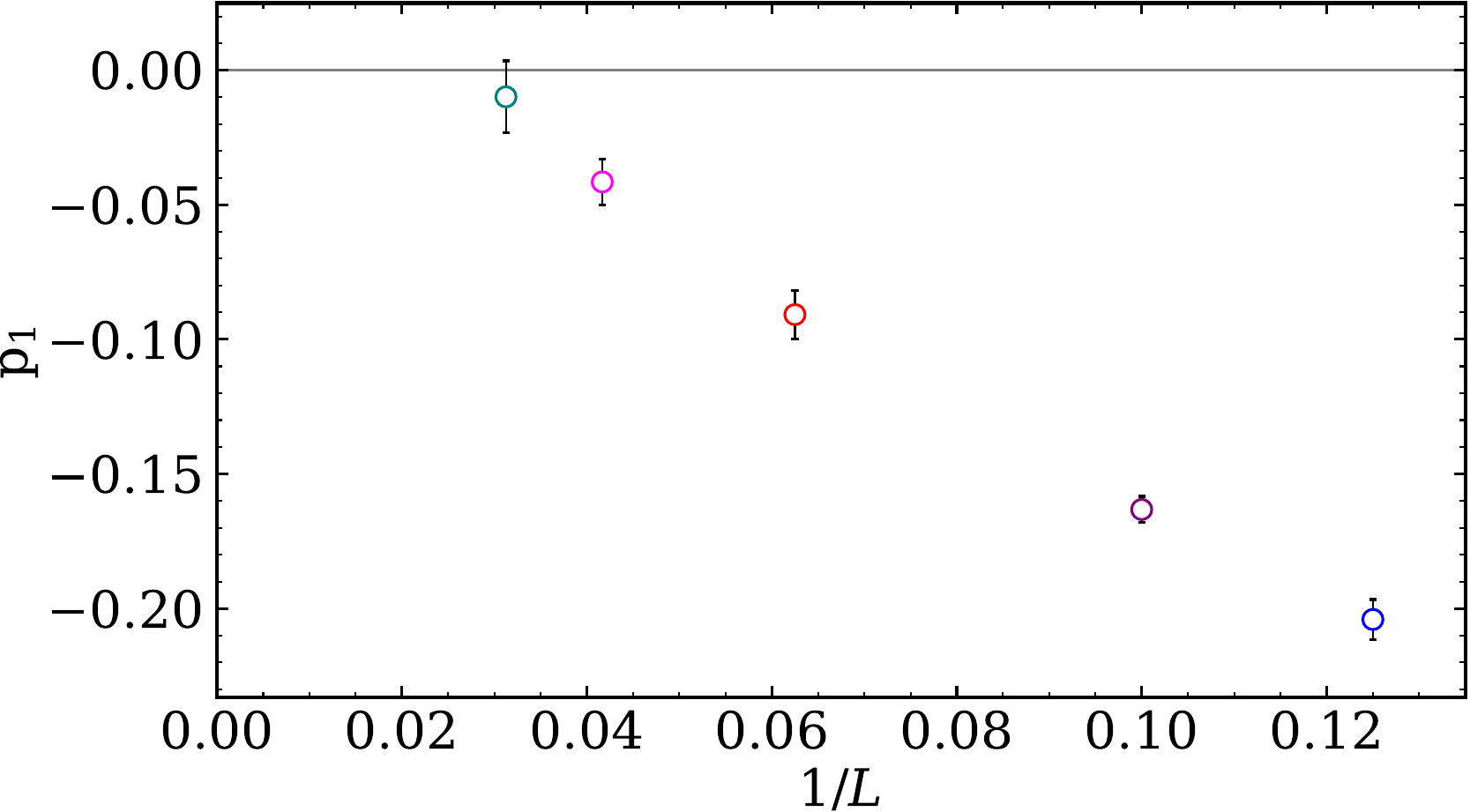}
    }    
    \vskip 0.00in
    \caption{Spectral densities at $W \!=\! 5$ in deep IR ($E < e^{-W/2}$) with 
    increasing $L$. Scales $1/L$ and $\Emin$ are marked by vertical lines. 
    Shaded bands delimit the regions $1\sigma$, $2\sigma$ and $3\sigma$ 
    away from $\Emin$ ($\sigma$ is the standard deviation). 
    Lower right: the power $p_1$ extracted from regions adjacent to 
    $\Emin + 3\sigma$, and shown by solid line segments, as a function 
    of $1/L$.
    }
    \label{fig:power_p1}
    \vskip -0.40in
\end{center}
\end{figure}


In this section we will focus on $p_1$. Before proceeding, we recall that 
in general
\begin{equation}
    p(E,W) \equiv \lim_{L \to \infty} \frac{d \log \rho(E,L,W)}{d \log E} 
    \qquad , \qquad E> 0  
    \label{eq:077} 
\end{equation}
so that the power is extracted from the slopes in log-log plots of DOS. 
Specifically then,
\begin{equation}
    p_1(W) \,\equiv\, \lim_{E \to 0^+} p(E,W) 
    \qquad , \qquad
    p_2(W) \,\equiv\, p(E\!=\!1, W) 
    \label{eq:078} 
\end{equation}
The task of extracting $p_1$ thus boils down to monitoring the slope 
of the density plots in the lowest part of the spectrum unaffected by 
finite-L effects. In order to probe where in the spectrum that is, 
we show in Fig.~\ref{fig:power_p1} density plots below $E \!=\! e^{-W/2}$ 
for a sequence of systems with increasing $L$ at $W \!=\!5$. One can 
observe the pattern of two spectral regimes at each $L$, roughly below 
and above $\Emin$. Since $\Emin$ goes rapidly to zero with increasing 
$L$ (see Sec.~4), the lower regime will be pushed to zero energy in 
the thermodynamic limit. Hence, it is the upper regime that defines $p_1$.


Notice that the transition regime around $\Emin$ has a finite width and it 
is thus necessary to adjust the lower edge of the stable spectral range 
accordingly. Denoting by $\sigma$ the standard deviation in 
the distribution of $E_{\scriptscriptstyle{\text{min}}}$ at given $L$, we 
found empirically that setting this lower edge to $\Emin + 3\sigma$ works 
well at this $W$. To see their extent, the bands for $1\sigma$, $2\sigma$ 
and $3\sigma$ away from $\Emin$ are shown in the plots 
of Fig.~\ref{fig:power_p1} via varying shades of grey. The boundaries 
of the spectral interval over which the limiting slope was estimated are 
related by the factor of $e$ and the corresponding linear segment is 
indicated for each $L$ in the sequence. The resulting slopes are shown 
in the lower right plot of Fig.~\ref{fig:power_p1}, and reflect the visually
compelling leveling-off the computed $\rho(E \to 0^+)$. 

Based on the above, we conclude that  
\begin{equation}
    p_1(W) \, = \, 0 \quad\;\; \text{for}  \quad\;\; W < W_0 
    \label{eq:088}
\end{equation}
where $W_0 \!>\! 5$ and may be infinite. In other words, DOS does 
{\em not} develop a power-law singularity at $E\!=\!0$ for such $W$. 

\begin{figure}[t]
\begin{center}
    \vskip -0.06in
    \centerline{
    \hskip -0.02in
    \includegraphics[height=4.4truecm,angle=0]{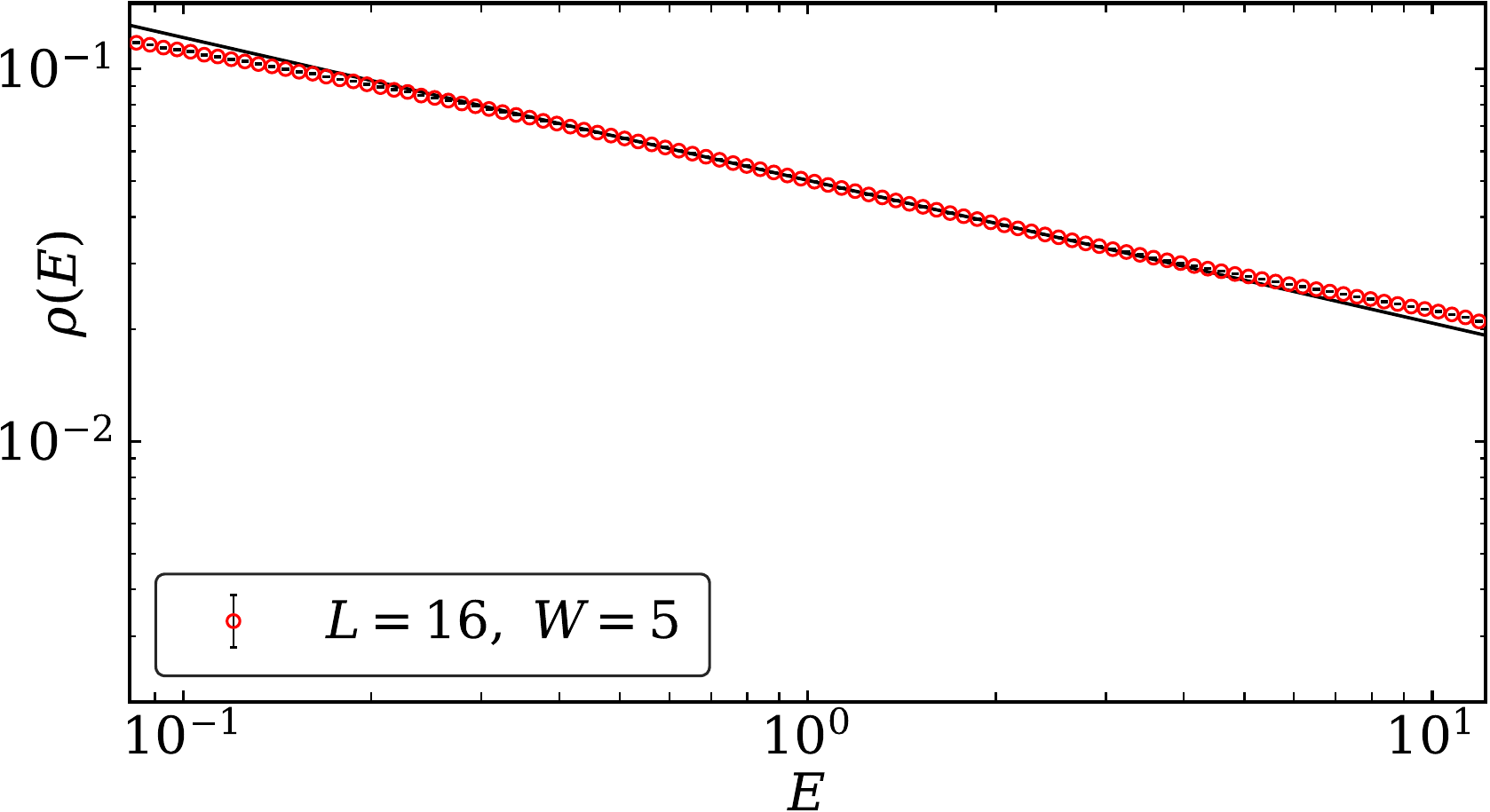}        
    \hskip 0.13in
    \includegraphics[height=4.4truecm,angle=0]{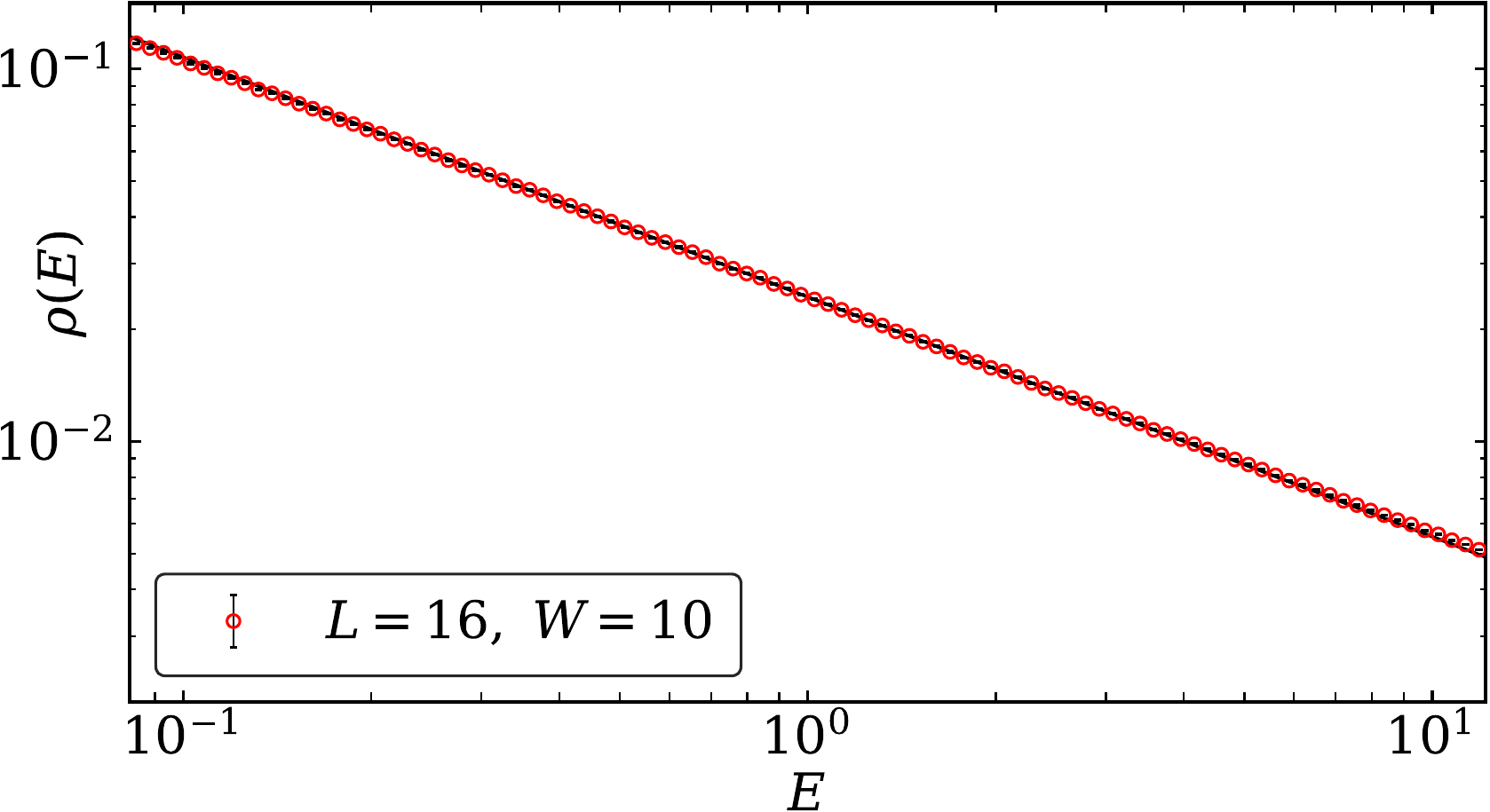}            
    }
    \vskip 0.13in
    \centerline{
    \hskip -0.02in
    \includegraphics[height=4.4truecm,angle=0]{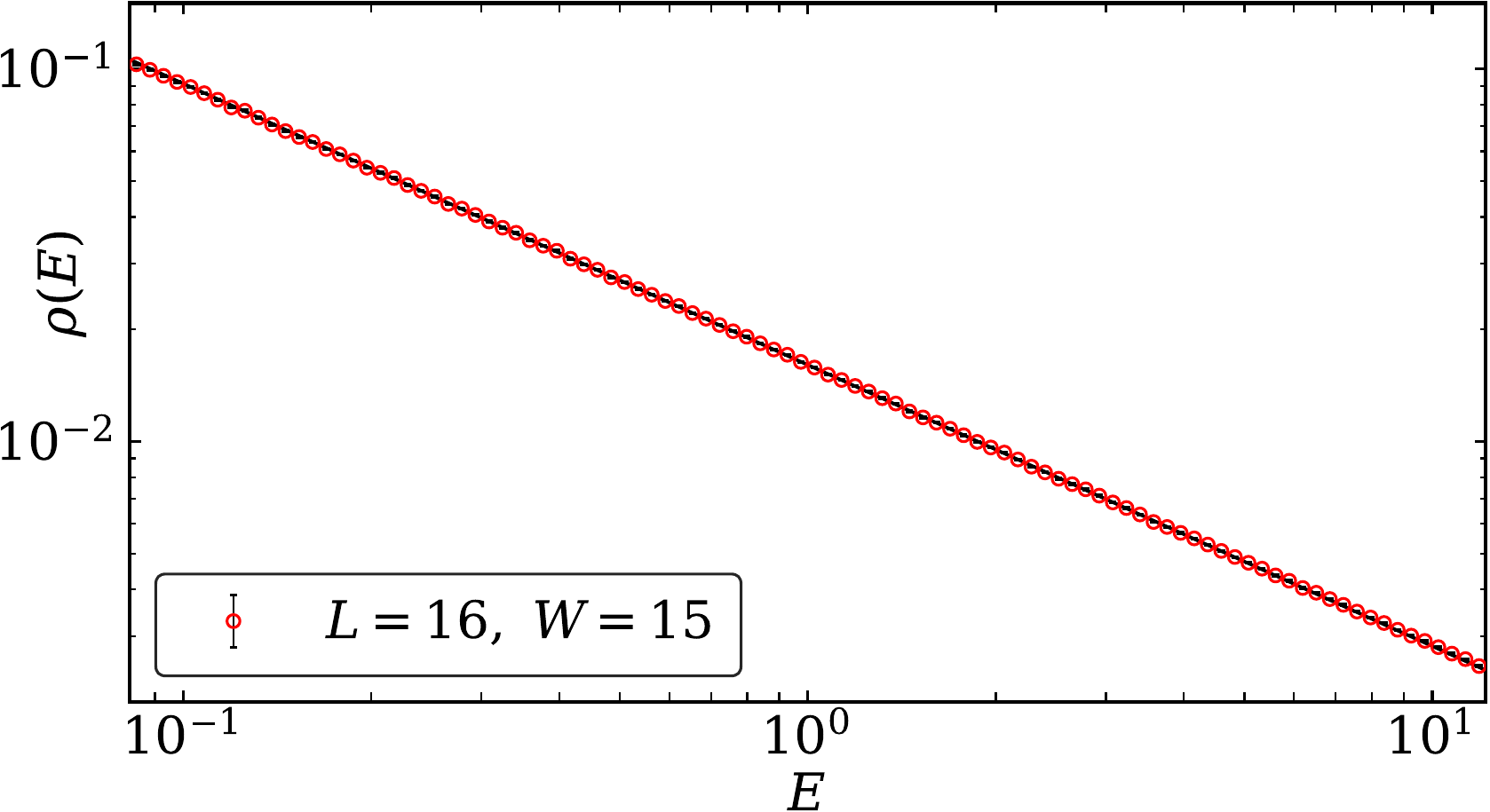}      
    \hskip 0.13in
    \includegraphics[height=4.4truecm,angle=0]{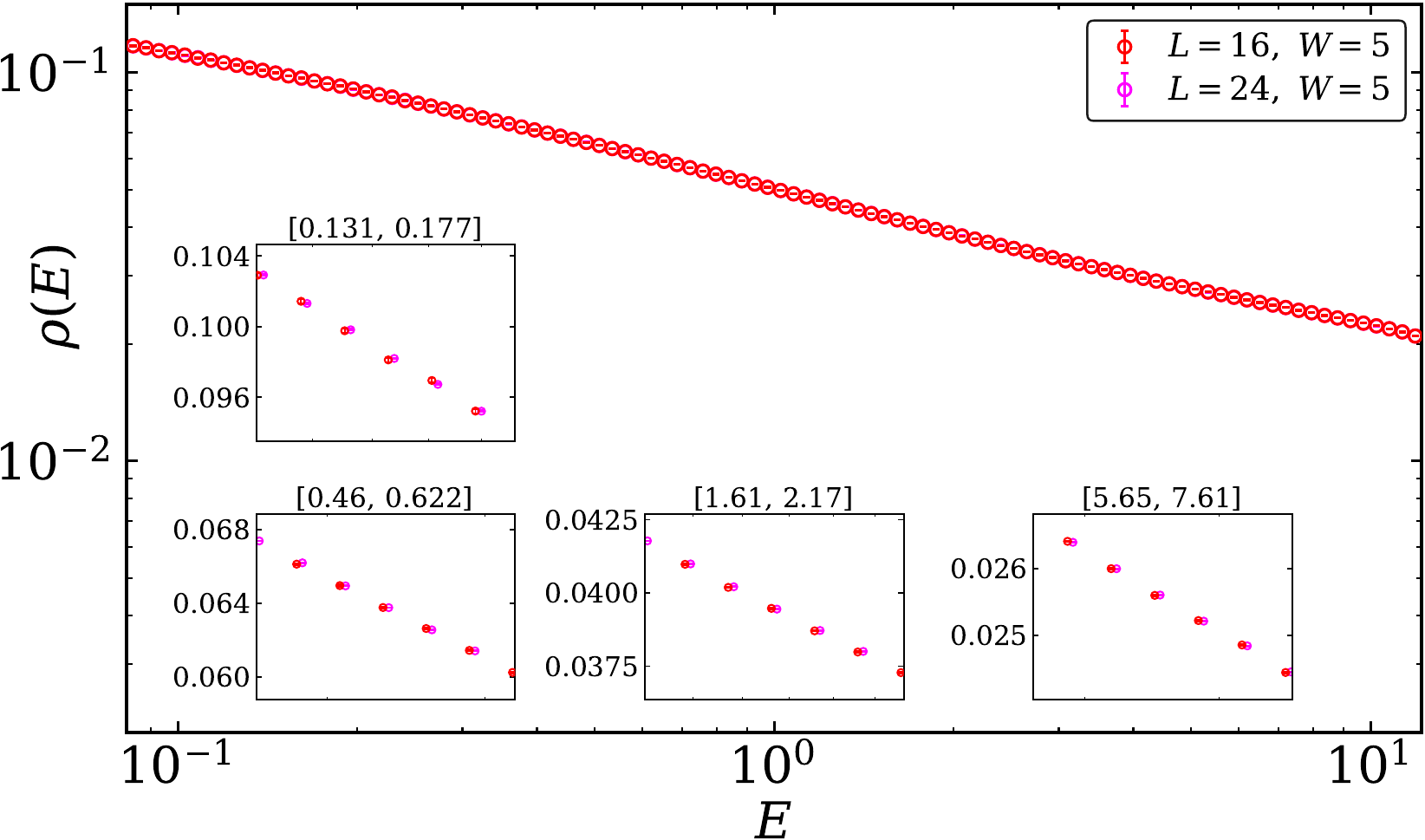}
    }    
    \vskip 0.00in
    \caption{
    Top left: log-log plot of $\rho(E)$ at $W \!=\! 5$ ($L\!=\!16$) in its noise cutoff range 
    $E \in [e^{-5/2}, e^{5/2}]$.       
    Top right \& bottom left: $\rho(E)$ in the same region but for systems at $W \!=\! 10$ and 
    $W \!=\! 15$ respectively. Bottom right: comparison of $L \!=\! 16$ and $L \!=\! 24$ at 
    $W \!=\!5$ (see the text).
    }
    \label{fig:density_around_1}
    \vskip -0.40in
\end{center}
\end{figure}

\smallskip
\noindent
{\bf 6.~Removal of the Noise Cutoff.}
We now focus on the behavior of DOS within the noise-cutoff range 
$[e^{-W/2}, e^{W/2}]$ which is of prime interest to us since,
as we found, this is how a true power-law singularity could develop
at $E \!=\! 0$ upon removing the noise cutoff ($W \to \infty$). In order 
to appreciate how accurate the power law is in this region, we show 
in Fig.~\ref{fig:density_around_1} this DOS at $W \!=\! 5$. We also 
included the straight line associated with the power $p_2$ extracted 
from the interval $[k,1/k]$ with $k\!=\!0.4$, i.e. from 
the interval $E \in [0.4,2.5]$. We see that perceivable deviations 
from pure power law only occur at the very edges of the full noise 
cutoff range. However, these violations effectively disappear in this 
fixed range, when we increase $W$ (top right and bottom left plot).
Note that the fits were done in the same range and that the slope
becomes steeper ($p_2$ more negative) with increasing $W$. 

Data in Fig.~\ref{fig:density_around_1}, and in what follows, is at 
$L \!=\! 16$ which is sufficient for this purpose (Sec.~4). To make
a targeted check, in Fig.~\ref{fig:density_around_1} (bottom right) 
we compare densities in the noise-regularization window for 
$L\!=\!16$ and $L \!=\! 24$ ($W \!=\! 5$). Four different spectral 
ranges are zoomed in by the insets, showing that the two do indeed 
agree at the level of statistical fluctuations.

\begin{figure}[t]
\begin{center}
    \vskip 0.05in
    \centerline{
    \hskip -0.02in
    \includegraphics[height=4.5truecm,angle=0]{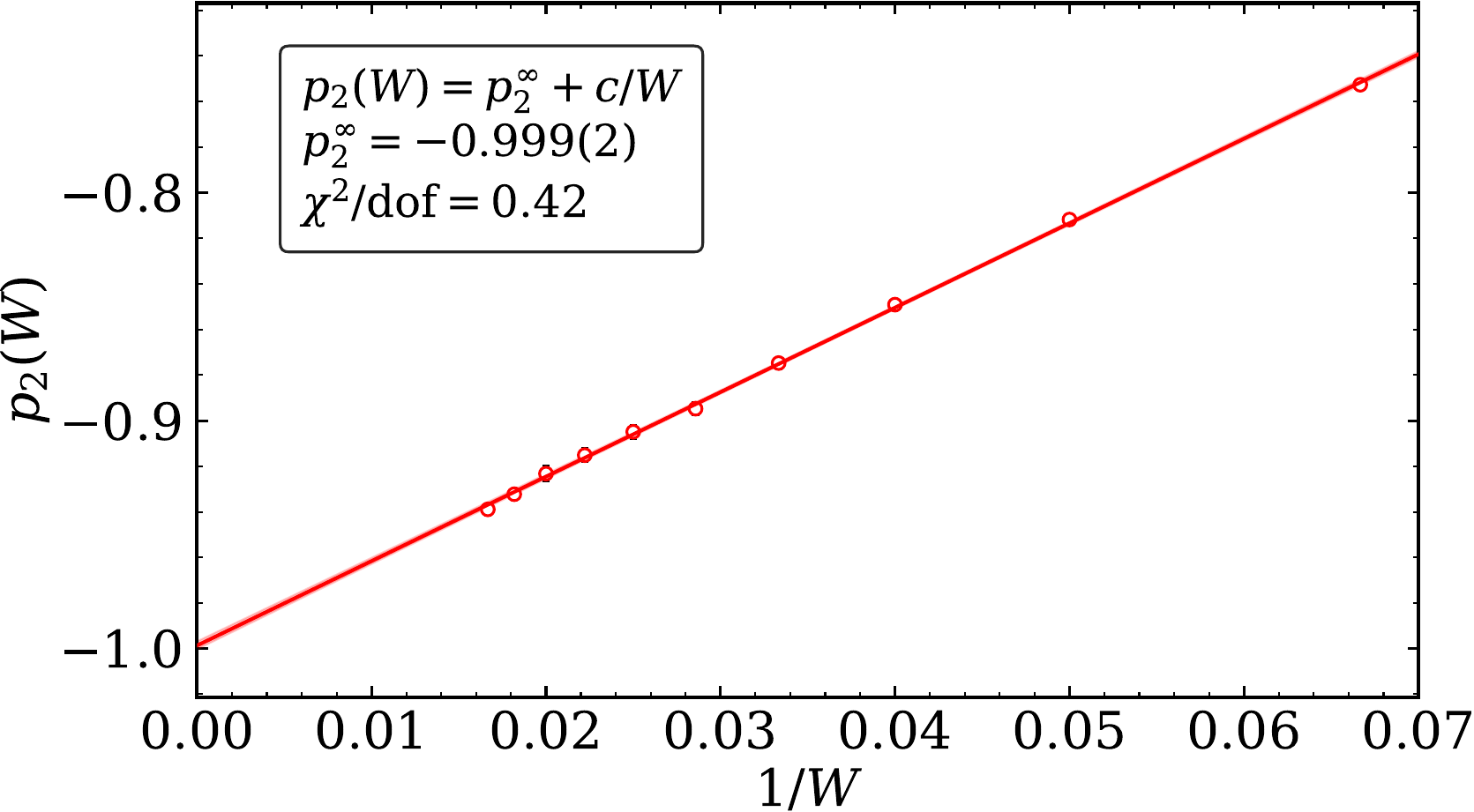}        
    \hskip -0.07in
    \includegraphics[height=4.5truecm,angle=0]{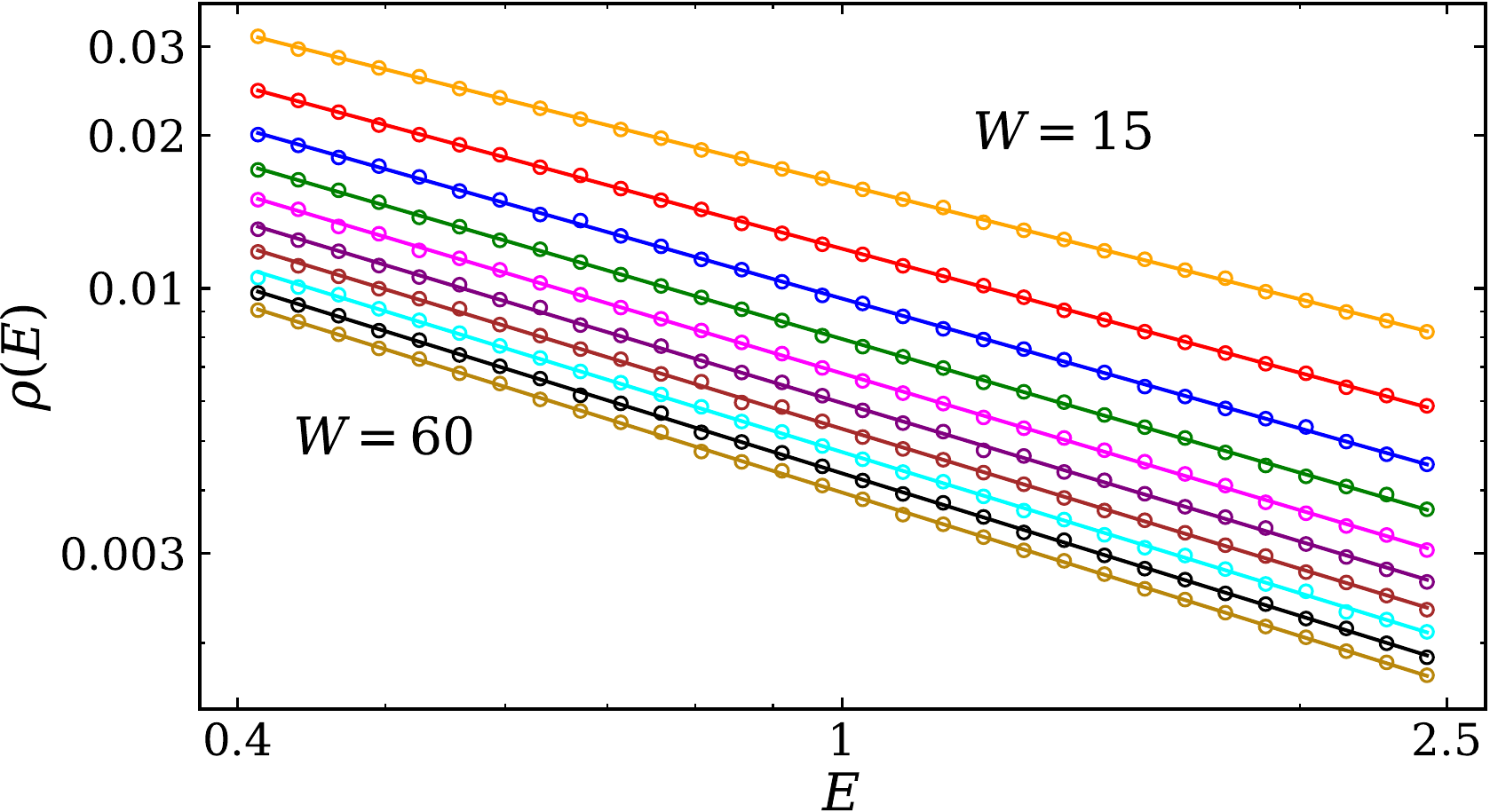}            
    }
    \vskip 0.00in
    \caption{Left: removing the noise regularization ($W \!\to\! \infty$ ) for $p_2$. 
    Right: log-log plot of densities at individual values of W over the common 
    fit range. 
    }
    \label{fig:remove_W}
    \vskip -0.40in
\end{center}
\end{figure}

In order to extract the cutoff-removed value of $p_2$, namely 
$\lim_{W \to \infty} p_2(W)$, we repeated the calculation, 
described above for $W\!=\!5, 10, 15$, at higher 
values of the cutoff, namely $W\!=\! 20,25,30,35,\ldots,60$. 
The results are shown in Fig.~\ref{fig:remove_W} (left) together 
with the associated plots of log-log spectral densities in the fitting 
range (right), demonstrating exceptionally precise power laws. 
The $p_2(W)$ data was fit to the constant with $1/W$ correction
and its parameters are shown. Based on this, we conclude that, 
upon removing the noise cutoff, we have
\begin{equation}
    p_2  \,\equiv\, \lim_{\Wdis \to \infty} \,p_2(\Wdis)  \; = \; -1
    \label{eq:098}
\end{equation}

\begin{figure}[t]
\begin{center}
    \vskip -0.06in
    \centerline{
    \hskip -0.02in
    \includegraphics[height=4.4truecm,angle=0]{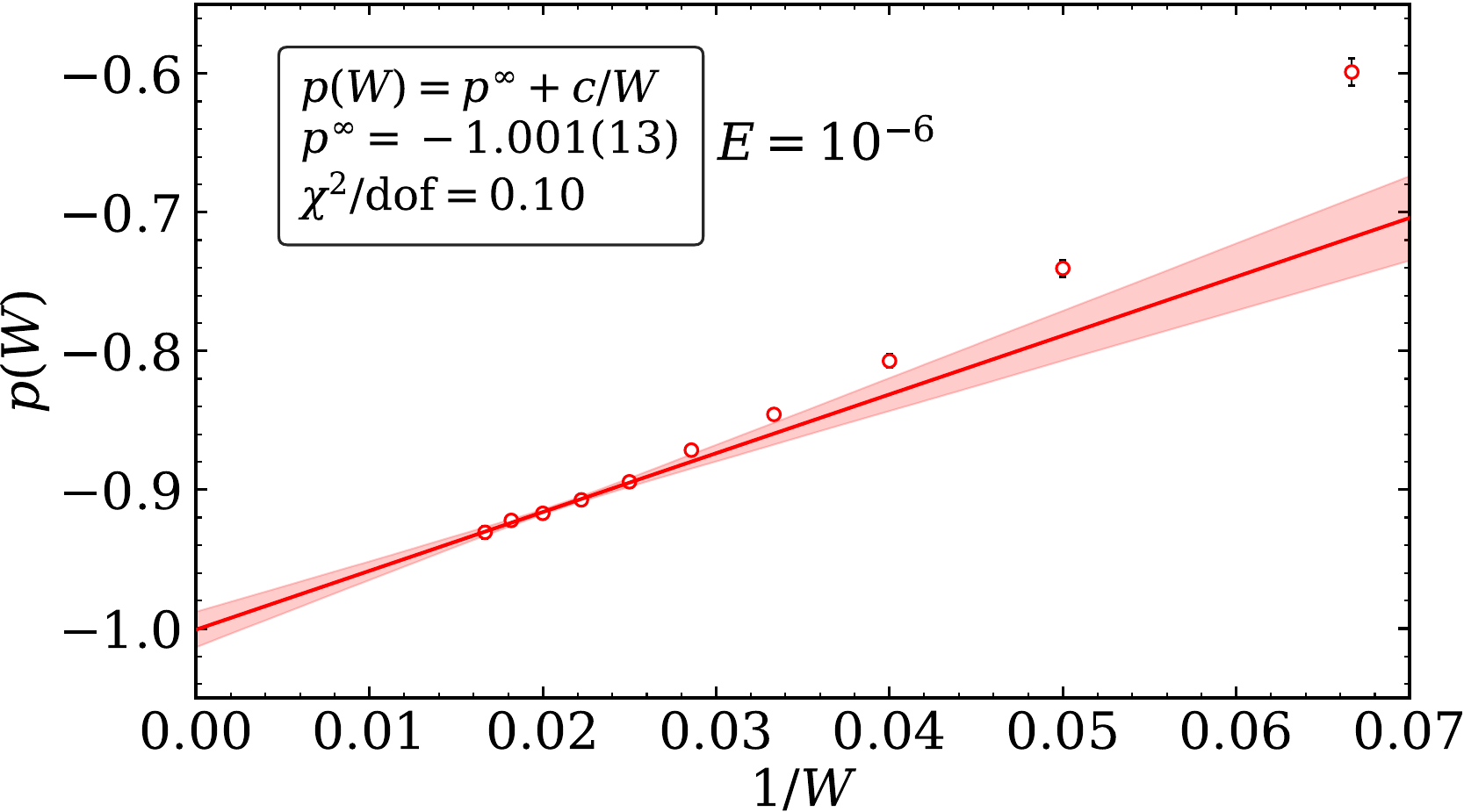}        
    \hskip 0.13in
    \includegraphics[height=4.4truecm,angle=0]{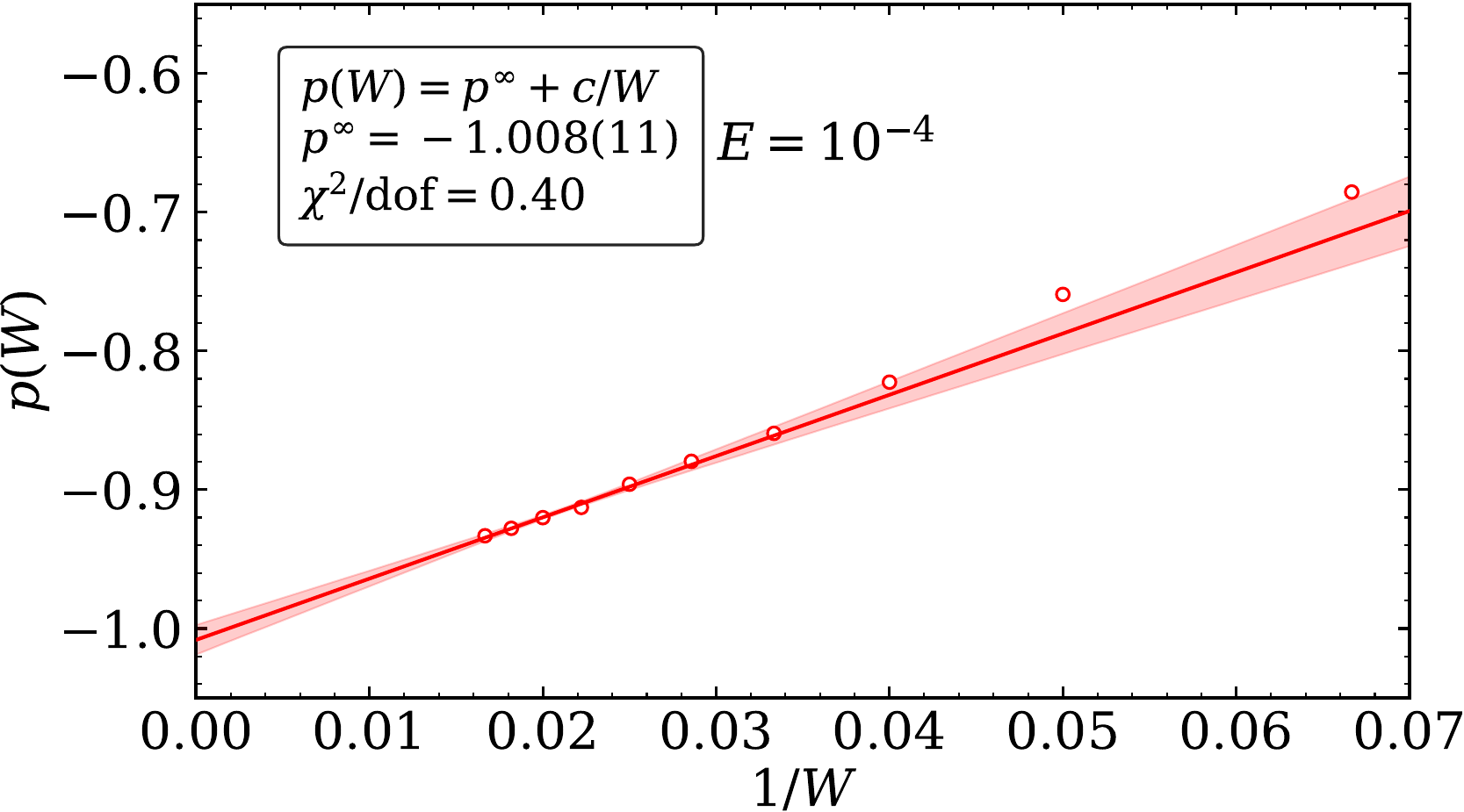}            
    }
    \vskip 0.13in
    \centerline{
    \hskip -0.02in
    \includegraphics[height=4.4truecm,angle=0]{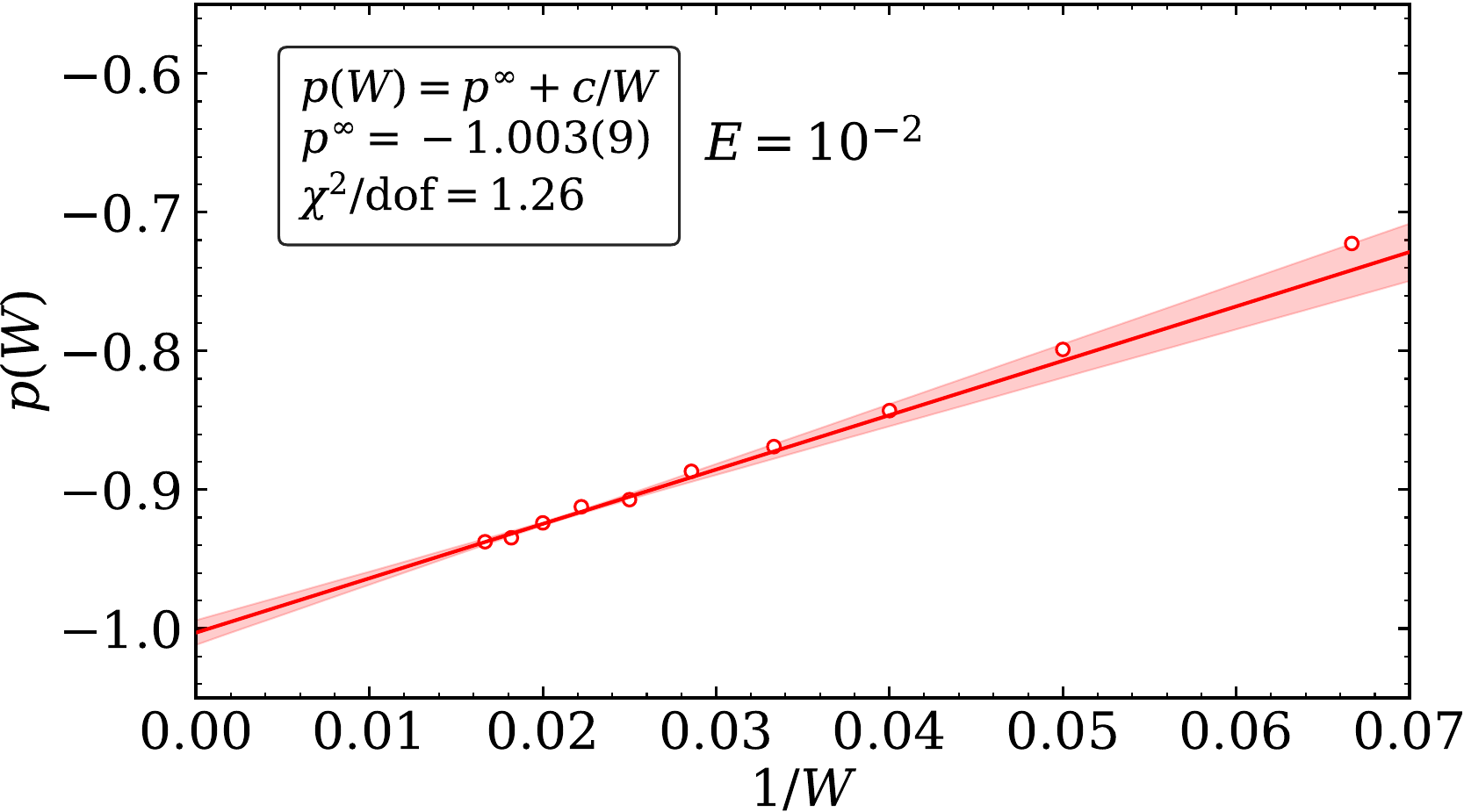}        
    \hskip 0.13in
    \includegraphics[height=4.4truecm,angle=0]{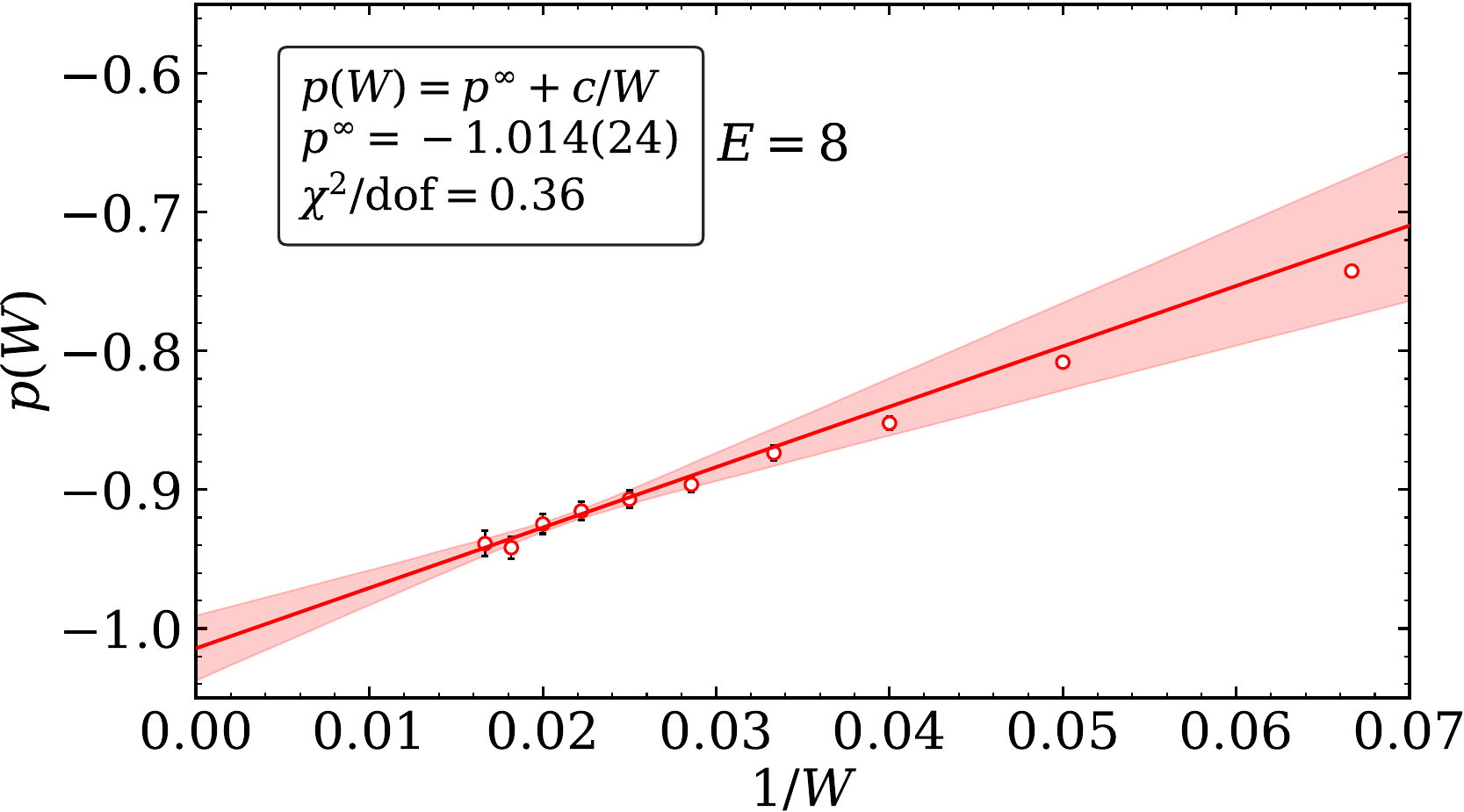}
    }    
    \vskip 0.00in
    \caption{
    Noise cutoff removal for systems at various energies. In all cases the fits
    were performed for $W \ge 40$ in order to avoid finite-$W$ effects. 
    See the discussion in the text.
    }
    \label{fig:density_around_E}
    \vskip -0.40in
\end{center}
\end{figure}

The plots in Fig.~\ref{fig:density_around_1} reveal that that the power law 
within the noise-cutoff interval $[e^{-W/2}, e^{W/2}]$ is very precise and 
spreads almost toward its edges. Hence, in the $W \to \infty$ limit, a single 
power law, is expected to govern the entire non-negative axis of 
the spectrum. To test this, we have computed the analogs of $p_2$, 
namely $p(E,W)$ for values of $E$ other than 1. 
In Fig.~\ref{fig:density_around_E} we show the results and $W \to \infty$ 
extrapolations for $E \!=\! 10^{-6}, 10^{-4} , 10^{-2}, 8$.
This was done in the same way as for $p_2$, namely by fitting pure power 
over the interval $[E k, E/k]$, using  $k \!=\! 0.4$ for all $E < 1$ and $k=2/3$
for $E \!=\! 8$.\footnote{The ratio for $E\!=\!8$ was
reduced simply because we did not cover the sufficiently large part of the 
spectrum in our eigenmode calculations in that case. We checked for results 
being very stable with respect to the choice of $k$.}

Notice that, e.g. for $E\!=\!10^{-6}$, there are $W$-s considered for which 
this energy is outside the noise cutoff region (below $e^{-W/2}$) or too
close to it. One thus expects violations for such $W$-s, and the fitting 
range has to be adjusted to get reliable results. This is clearly visible in 
the plots of Fig.~\ref{fig:density_around_E}. To avoid such contamination,
we use the common fitting range $W \!\ge 40$ for all these energies.
Our results strongly suggest that 
\begin{equation}
    p(E) \equiv \lim_{W \to \infty} p(E,W) = -1  \quad \text{for all} \quad E > 0
    \label{eq:108}
 \end{equation} 
Moreover, the behavior of $\rho(E)$ within the noise cutoff bounds (see e.g.
Fig.~\ref{fig:density_around_1}) also strongly suggests that 
the convergence to $p(E) \equiv -1$ is uniform on any fixed positive-energy 
interval $(E_1, E_2)$.

\vfill\eject

\smallskip
\noindent
{\bf 7.~Discussion and Outlook.}
In this work we have shown, via systematic numerical investigation, that
quantum mechanics transforms the scale-invariant $1/t$ noise of random 
hoppings~\eqref{eq:048}, into a scale-invariant $1/E$ distribution 
of eigenstates. We were seeking this because such behavior exactly 
mimics the interpretation of the scale-invariant Dirac spectral density of 
the QCD IR phase as arising from self-similar (scale-invariant) glue 
backgrounds~\cite{Alexandru:2019gdm, Alexandru:2021pap, Alexandru:2021xoi}.

The details and the meaning of the above statement are important, and 
they have to do with the necessity to regularize the $1/t$ noise (in fact 
any power-law noise) on the non-negative real axis. Chiral-orthogonal 
Anderson models with off-diagonal logarithmic disorder realize this cutoff 
via their randomness strength $W$ and the full scale invariance is only 
realized in the $W  \!\to\! \infty$ limit. From our point of view, 
the $W$-family of these models is a single model with cutoff: 
the regulator of scale invariance.

While our results can be viewed as a model proof of concept for 
the proposed realization of scale invariance in QCD IR phase, 
interesting questions arise if one aims to use this acquired knowledge
to actually model the phase (e.g. via matching certain properties of 
eigenmodes).  Although the full class of related relevant models is yet 
to be defined, what stands out in our present results is that the lower
spectral range for scale invariance in the model is chiefly controlled
by the noise-cutoff scale $e^{-W/2}$ rather than scale related to $1/L$, 
as observed so far in QCD~\cite{Alexandru:2019gdm}. The two will 
thus need to be put into a suitable correspondence for such modeling 
to succeed.

We also note in closing that other type of models, namely mixing 
models involving unit topological lumps, have previously been studied 
in this context~\cite{Edwards:1999zm, Kovacs:2023vzi}. 
The difference in our approach is that it focuses on the scale invariance 
aspect directly.\footnote{It also literally follows the simplest version of 
metal-to-critical scenario~\cite{Alexandru:2021xoi} with the IR part
described by the ordinary Anderson model.}
Further research will show whether certain combination of both 
approaches may supply the idea for most faithful model representation.

\vfill\eject

\noindent
{\bf Acknowledgments.}
We thank Peter Marko\v{s} and Robert Mendris for discussions related
to the subject of the paper. This work was supported by the U.S. Department 
of Energy under Grant~No.~DE-FG02-95ER40907, by the Department of Atomic 
Energy, Government of India, under Project Identification Numbers RTI-4012 
and RTI-4013, and by the Nuclear Physics Institute CAS. Computations were 
carried out, in part, on the computing clusters at the Department 
of Theoretical Physics, TIFR, Mumbai and also on MetaCentrum servers 
provided by the e-INFRA CZ project (ID:90254), supported by the Ministry 
of Education, Youth and Sports of the Czech Republic.

\bibliographystyle{JHEP} 
\bibliography{chiralanderson}

\medskip
\noindent
{\bf A.~FEAST Eigensolver.}
We compute the eigenvalues and eigenvectors of the disordered hopping
Hamiltonian of Eq.~\eqref{eq:018} using the \feast{}
eigensolver~\cite{polizzi2009density, 7877094, di2016efficient, gavin2017improved, polizzi2020feast}. 
\feast{} returns all eigenvalues in a chosen energy interval, together with
their eigenvectors. This eigensolver has previously been applied in other 
fields, but to our knowledge, this is its first application in the context 
of chiral Anderson models, which are motivated by the lattice QCD Dirac operator.
In this appendix, we describe the method and how we use
it for our purpose.

For this work, we consider a cubic lattice of $L\times L\times L$ sites with periodic
boundary conditions. The Hamiltonian is a real symmetric $N\times N$ matrix,
with $N=L^3$, containing only nearest-neighbor hopping terms and no on-site
term. Each site has six nearest neighbors, so each row of $H$ contains six
non-zero entries. The total number of non-zero entries is therefore $6N$,
compared with $N^2$ for a dense matrix. We store $H$ in compressed sparse
row (CSR) format and never construct the dense matrix.

The \feast{} eigensolver selects the eigenvalues in a window as follows. Consider an energy
interval $\mathcal{I}=[E_{\min}, E_{\max}]$ and a closed contour
$\mathcal{C}$ in the complex plane that encloses $\mathcal{I}$ and no other
eigenvalues of $H$. The projector onto the eigenstates with energies in
$\mathcal{I}$ is a contour integral of the resolvent,
\begin{equation}
P=\sum_{E_n\in\mathcal{I}}|\psi_n\rangle\langle\psi_n|
 =\frac{1}{2\pi i}\oint_{\mathcal{C}}dz\,(z-H)^{-1}.
\label{eq:projector_app}
\end{equation}
This follows from Cauchy's theorem: each pole $z=E_n$ inside $\mathcal{C}$
contributes $|\psi_n\rangle\langle\psi_n|$, while poles outside do not contribute.

The \feast{} evaluates Eq.~\eqref{eq:projector_app} by numerical quadrature,
\begin{equation}
P\approx\sum_{k=1}^{N_c}w_k\,(z_k-H)^{-1},
\label{eq:quadrature_app}
\end{equation}
with nodes $z_k$ on $\mathcal{C}$ and weights $w_k$. The algorithm starts
from a block $Y$ of $M_0$ random vectors, where $M_0$ is larger than the
number $M$ of eigenvalues in $\mathcal{I}$. It first computes $Q=PY$ with
Eq.~\eqref{eq:quadrature_app}. This requires one sparse linear solve,
$(z_k-H)X_k=Y$, per node, each with $M_0$ right-hand sides. Note that the matrix $H$
is never inverted or diagonalised. It then projects $H$ onto the space
spanned by $Q$ and solves the small $M_0\times M_0$ problem
$(Q^{T}HQ)\,v=E\,(Q^{T}Q)\,v$. This Rayleigh--Ritz step gives approximate
eigenvalues $E_n$ and eigenvectors $\psi_n=Qv_n$. The approximate
eigenvectors are used as the new $Y$, and the two steps are repeated until
the residuals are small. \feast{} is therefore a subspace iteration in which
the projector of Eq.~\eqref{eq:projector_app} selects the
window.

To see the role of the quadrature, one applies the sum in
Eq.~\eqref{eq:quadrature_app} to a single eigenvector $|\psi_n\rangle$. The
result is $f(E_n)\,|\psi_n\rangle$, with
\begin{equation}
f(E)=\sum_{k=1}^{N_c}\frac{w_k}{z_k-E}.
\label{eq:filter_app}
\end{equation}
The exact integral gives $f=1$ for $E$ inside $\mathcal{I}$ and $f=0$
outside. With a finite number of nodes, $f(E)$ is close to 1 inside the
window, but close to 0 well outside it, and drops smoothly near $E_{\min}$ and
$E_{\max}$. The quadrature therefore acts as a filter in energy: it keeps
the states in the window and suppresses all others. The suppression is not
perfect, so a few iterations are needed, and each one further reduces the
weight left outside the window. More nodes give a sharper filter and faster
convergence, at the cost of more linear solves per iteration.

Since $H$ is real symmetric, its eigenvalues are real, and the contour can
be chosen symmetric about the real axis. The contributions from the lower
half of the contour are then the complex conjugates of those from the upper
half, so only the nodes on the upper half need to be computed. This halves
the number of linear solves.

The size of the search subspace is set by the number of eigenvalues
expected in the window, not directly by the matrix dimension $N$. Degenerate
eigenvalues are obtained together with their complete set of eigenvectors.
Different windows are independent calculations, so a larger part of the
spectrum can be covered by adjacent windows, computed one after another or
in parallel. This allows us to scan the spectrum and spend computing time
only on the regions of interest.

We use the sparse real-symmetric driver \texttt{dfeast\_scsrev} of
\feast{}~\cite{polizzi2020feast}, with $H$ stored in CSR format. We solve the linear
systems $(z_k-H)X_k=Y$ in double precision with the sparse direct
solver MKL-PARDISO, and the quadrature uses Gauss--Legendre nodes on a
half-ellipse above the real axis. A calculation is converged when every
eigenpair in the window has a relative residual below a fixed tolerance. To
reduce memory use, we do not keep matrix factorisations between iterations. 
Since the lattices considered here are bipartite and $H$ has no on-site
term, the spectrum is symmetric about $E=0$, so we place the window at
non-negative energies. For each $(L,W)$, we generate at least $10^4$
independent disorder realisations and run \feast{} on each. 

The same eigenpairs can be obtained with other methods, such as full
diagonalisation for small lattices or shift-invert Lanczos. Full
diagonalisation computes the whole spectrum, at a cost of
$\mathcal{O}(N^3)$ operations and $\mathcal{O}(N^2)$ memory, while our
analysis needs only part of it. Shift-invert Lanczos uses the sparsity of
$H$, but it finds the eigenvalues closest to a chosen shift, so covering a
full interval requires several shifts and a check that no eigenvalue has
been missed. \feast{} instead returns all eigenvalues in a prescribed
interval, together with their eigenvectors, using only sparse linear
solves. The eigenvectors in the window also give access to observables
beyond the spectral quantities considered in this work. These will be
studied in a forthcoming publication.

\end{document}